\documentclass[journal]{IEEEtran}
\usepackage{cite}

\ifCLASSINFOpdf
\usepackage[pdftex]{graphicx}
\else
\fi
\pdfoutput=1

\usepackage{amsmath}
\usepackage{amssymb}
\usepackage{eurosym}
\usepackage{romannum}

\usepackage{algorithmic}

\usepackage{array}
\usepackage{booktabs}

\usepackage{xcolor}
\definecolor{rev1}{RGB}{0,0,0}
\definecolor{rev2}{RGB}{0,0,0}

\begin{document}
%
\title{Convex Relaxations and Approximations of Chance-Constrained AC-OPF Problems}
%
%
%
\author{Lejla~Halilba\v{s}i\'{c},~\IEEEmembership{Student Member,~IEEE,}
        Pierre~Pinson,~\IEEEmembership{Senior Member,~IEEE,}
        and~Spyros~Chatzivasileiadis,~\IEEEmembership{Senior Member,~IEEE}
\thanks{This work is supported by the EU project BEST PATHS, Grant No. 612748.}
\thanks{L. Halilba\v{s}i\'{c}, P. Pinson, and S. Chatzivasileiadis are with the Department
of Electrical Engineering, Technical University of Denmark, Kongens Lyngby, Denmark. email: \{lhal, ppin, spchatz\}@elektro.dtu.dk}
}
%
%
\markboth{ACCEPTED FOR PUBLICATION IN IEEE TRANSACTIONS ON POWER SYSTEMS}%
{Shell \MakeLowercase{\textit{et al.}}: Bare Demo of IEEEtran.cls for IEEE Journals}
%
\maketitle
\begin{abstract}
This paper deals with the impact of linear approximations for the unknown nonconvex confidence region of chance-constrained AC optimal power flow problems. Such approximations are required for the formulation of tractable chance constraints. In this context, we introduce the first formulation of a chance-constrained second-order cone (SOC) OPF. The proposed formulation provides convergence guarantees due to its convexity, while it demonstrates high computational efficiency. Combined with an AC feasibility recovery, it is able to identify better solutions than chance-constrained nonconvex AC-OPF formulations. To the best of our knowledge, this paper is the first to perform a rigorous analysis of the AC feasibility recovery procedures for robust SOC-OPF problems. We identify the issues that arise from the linear approximations, and by using a reformulation of the quadratic chance constraints, we introduce new parameters able to reshape the approximation of the confidence region. We demonstrate our method on the IEEE 118-bus system.
\end{abstract}

%
\IEEEpeerreviewmaketitle

\section{Introduction}
%
%
%
%
\IEEEPARstart{P}{ower} system operations increasingly rely on the AC Optimal Power Flow (OPF) to identify optimal decisions \cite{opf_basic_requirements}, while
higher shares of intermittent renewable generation add an additional layer of complexity and call for modeling approaches which account for uncertainty. Literature considers uncertainty either in the form of stochastic formulations, which optimize over several possible realizations (i.e. scenario-based), or in the form of robust formulations, where chance constraints are incorporated in the optimization problem accounting for a continuous range of uncertainty. This paper focuses on chance-constrained optimization.

\begin{figure}
\centering
\includegraphics[scale=0.2]{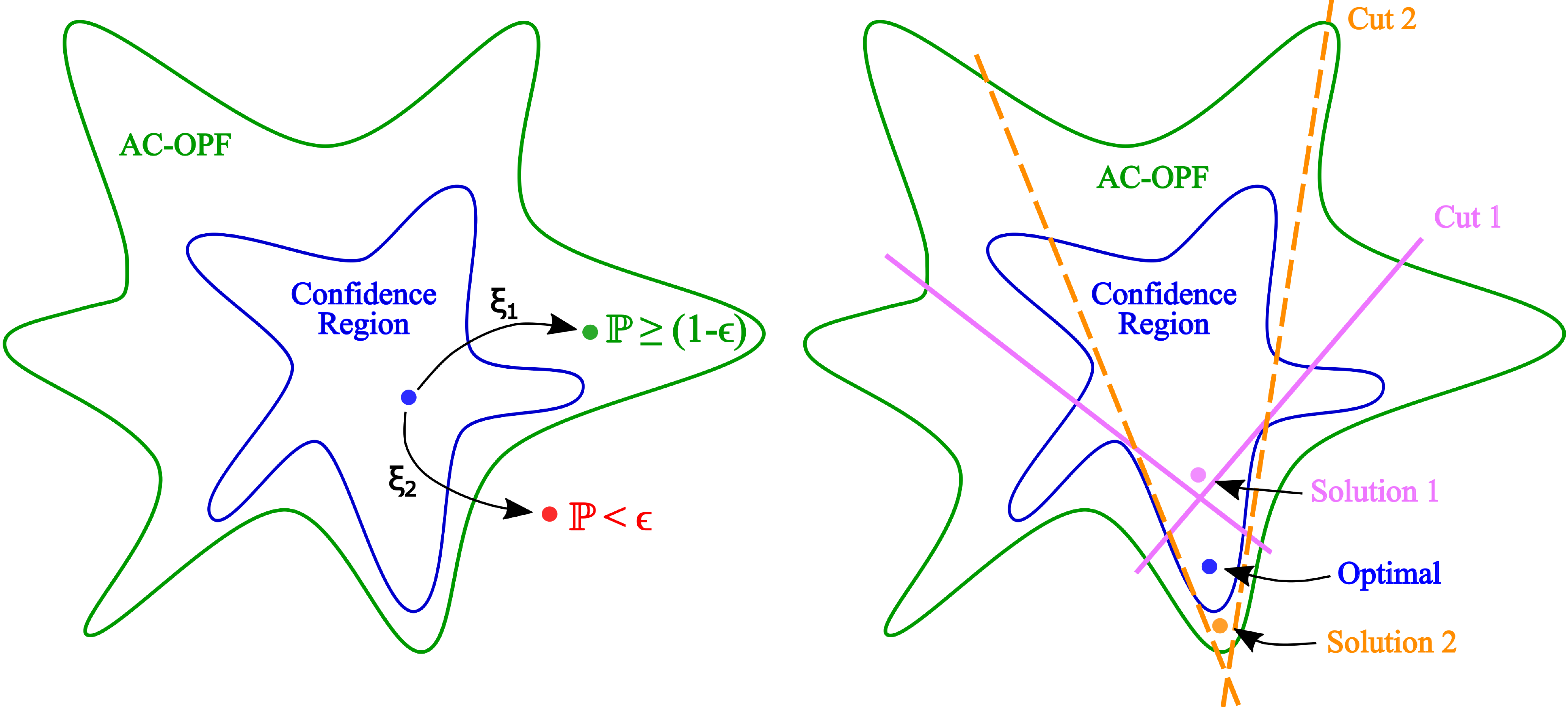}
\centering \caption{Left: Illustration of the feasible space of the AC-OPF and the chance-constrained AC-OPF (confidence region). Right: Illustration of the approximation of the confidence region using linear cuts.}
\vspace{-15pt}
\label{fig:FeasibleSpace}
\end{figure}

Chance constraints define the maximum allowable violation probability $\epsilon$ of inequality constraints and reduce the nonconvex feasible space of the AC-OPF to a desired confidence region, which is also nonconvex as depicted in blue in Fig. \ref{fig:FeasibleSpace}. This confidence region includes only operating points which under any realization of the uncertainty $\xi$ are guaranteed to remain within the feasible space of the original AC-OPF (in green in Fig.~\ref{fig:FeasibleSpace}) with a probability of at least $(1-\epsilon)$. {\color{rev1}The notion of preventively securing the system against uncertainty by restricting the feasible space is also in line with the concept of transmission reliability margins used for the cross-border capacity management in the ENTSO-E region \cite{cwe_fbmc}. Additionally, chance constraints offer the benefit of being relatively easily adaptable to a wide range of uncertainty behavior and safety requirements. Their application ranges from robust formulations -- such as targeting joint chance constraints \cite{maria_probabilistic_scopf,Andreas_CCACOPF}, not relying on the assumption of any distribution \cite{Baker_DistributionAgnosticOPF,Line_DistributionallyRobustSCOPF} or accounting for a family of possible distributions (i.e., distributionally robust) \cite{DR_ACOPF,DistributionallyRobust_DCOPF,Line_DistributionallyRobustSCOPF,2018arXiv180406388G} -- to less conservative frameworks, which consider single system constraints with different levels of robustness \cite{Line_AnalyticalReformulation,Line_CCACOPF}. The latter accounts for the fact that there are usually only few active constraints \cite{Bienstock_RiskAwareNetworkCtrl}, which could compromise system security and need to be treated more cautiously.}

As the AC-OPF is a nonlinear and nonconvex problem, it is impossible to formulate tractable chance constraints able to cover the whole continuous uncertainty space. Instead, literature has proposed tractable approximations.
{\color{rev1}Recent research has focused on developing formulations of the chance-constrained AC-OPF based on either partial or full linearizations, and analytical chance constraint reformulations \cite{Jeremias_CCACOPF,Line_CCACOPF,LineDan_IREP,FirstCCOPF,miles_yuri_line,Baker_DistributionAgnosticOPF}, while other works propose a combination of convex relaxations based on semidefinite programming (SDP) for the power flow equations and a scenario-based reformulation of the chance constraints \cite{maria_probabilistic_scopf,Andreas_CCACOPF}.}

The main challenge of the chance-constrained AC-OPF lies in approximating the unknown nonconvex confidence region. Common to all approaches in \cite{Line_DistributionallyRobustSCOPF,Line_AnalyticalReformulation,Line_CCACOPF,LineDan_IREP,Jeremias_CCACOPF,FirstCCOPF,maria_probabilistic_scopf,Andreas_CCACOPF,miles_yuri_line,Baker_DistributionAgnosticOPF,DR_ACOPF,DistributionallyRobust_DCOPF,Bienstock_RiskAwareNetworkCtrl} is that they approximate the impact of the uncertainty by a linearization allowing to reformulate the chance constraints to tractable deterministic constraints. These are tighter than the original AC-OPF constraints and represent linear cuts to the original feasible space in order to approximate the confidence region. As visualized in Fig. \ref{fig:FeasibleSpace}, depending on the quality of the cuts the identified operating points may either lie outside the confidence region (solution 2) \cite{Line_CCACOPF} or are too conservative (solution 1) {\color{rev1}as in the case of sample-based reformulations \cite{maria_probabilistic_scopf,Andreas_CCACOPF} and the distributionally robust case in \cite{DR_ACOPF}. A less conservative distributionally robust OPF has recently been proposed in \cite{2018arXiv180406388G} which considers ambiguity sets of distributions based on historical forecast error data and the Wasserstein metric. Data-driven DRO frameworks are a promising intermediate approach between stochastic optimization which rely on the assumption of a certain distribution, and robust optimization for the worst-case uncertainty realization. They leverage the knowledge from observed historical data and other statistical information to provide robustness for the uncertainty distribution. However, several challenges remain such as the choice of an appropriate radius for the ambiguity set and maintaining computational efficiency under the necessary sample based reformulations.
} 

The authors in \cite{Line_CCACOPF,LineDan_IREP,Jeremias_CCACOPF} develop an iterative framework for approximating the chance-constrained AC-OPF by alternating between an AC-OPF and a computation of the constraint tightenings (i.e., the linear cuts) based on a first-order Taylor series expansion around the forecasted operating point{\color{rev1}, which is more accurate than the full linearization in \cite{Baker_DistributionAgnosticOPF}}. Due to the nonconvex nature of the AC-OPF, however, the algorithm is not guaranteed to converge. {\color{rev1}Convergence and robustness are still challenges even for the standard AC-OPF, which particularly for large networks often fails to succeed \cite{Anders_RobustnessAndScalabilityOfSDP,ferc_SolutionTechniquesACOPF}}. {\color{rev1}The authors in \cite{miles_yuri_line} use a first order Taylor expansion to linearize the AC power flow equations around the forecasted operating point and to model the uncertainty impact. The resulting approximation of the chance-constrained AC-OPF achieves a high computational efficiency due to its convexity and an improved cost performance by optimizing over affine response policies. Despite its increased robustness though, the method still relies on the availability of an AC-OPF solution at the forecasted operating point to allow for the linearization of the power flow equations. Otherwise, the method's solution quality is determined by the quality of the input AC-OPF solution, which can be highly suboptimal \cite{ferc_SolutionTechniquesACOPF}.} The SDP relaxation of the chance-constrained AC-OPF developed in \cite{maria_probabilistic_scopf} improves on the approximation of the confidence region by optimizing over affine control policies, while in \cite{Andreas_CCACOPF} we additionally aim at providing AC feasible solutions and global optimality guarantees. However, as SDP solvers are still under development it can be computationally challenging.

This paper focuses on second-order cone relaxations (SOC) of the AC-OPF as a good trade-off between approaches for two reasons. First, compared with the original AC-OPF formulation, SOC relaxations define a convex problem which is guaranteed to converge. Second, SOC relaxations are computationally more efficient than SDP relaxations. It must be noted that compared to SDP, SOC relaxations provide a less tight relaxation, and require strengthening \cite{StrongSOCP} or other procedures to recover an AC feasible point. Such procedures are often necessary in the SDP formulation as well though.

SOC-OPF algorithms considering uncertainty have been proposed in \cite{AdaptiveRobustMultiPeriodACOPF,LetterRobustACOPF,AAROPF_ACDC}, where the authors develop convex formulations of the robust two-stage AC-OPF problem focusing on the worst-case uncertainty realization. Specifically, in \cite{LetterRobustACOPF} and \cite{AAROPF_ACDC} SOC relaxations are used within the framework of an affinely adjustable robust OPF (first proposed in \cite{Jabr_AAROPF} for a DC-OPF). However, both papers consider only affine policies for active power generation neglecting the impact of the uncertainty on all other control and state variables. A very extensive framework for relaxations of robust AC-OPFs is provided in \cite{AdaptiveRobustMultiPeriodACOPF}, where the authors develop three methods using conic duality to obtain tractable formulations of the robust AC-OPF based on SOC-, SDP-, and DC-OPFs. To guarantee AC feasible solutions, the conic OPF models are used to approximate the second stage of the two-stage robust optimization problem and are then solved alternately with an AC-OPF, which represents the first stage problem. However, as in \cite{Line_CCACOPF} this results in a nonconvex iterative program, which is not guaranteed to converge. None of the papers mentioned address the issue of AC infeasibility of the SOC-OPF solutions.

The main contributions of this work are:
\begin{itemize}
    \item the first formulation of a chance-constrained SOC-OPF (CC-SOC-OPF), able to provide both convergence guarantees and high computational efficiency; coupled with an AC feasibility recovery it {\color{rev1}can} identify better solutions than the chance-constrained nonconvex AC-OPF formulation
    \item the approximation of quadratic apparent power flow chance constraints with linear chance constraints using results proposed in \cite{TwoSided_LinCC}
    \item the introduction of new parameters able to reshape the approximation of the confidence region, along with a rigorous analysis of the linear approximations; these parameters offer a high degree of flexibility for the robustness of the solution.
\end{itemize}

The remainder of this paper is organized as follows: Section \ref{sec:MethodSOC} introduces the {\color{rev1}approximation} of the AC-OPF {\color{rev1}based on a SOC relaxation}, while Section \ref{sec:POPF} focuses on the formulation of the CC-SOC-OPF. Results from a case study are presented in Section \ref{sec:CaseStudy}. Section \ref{sec:Conclusion} concludes and Section \ref{sec:FutureWork} discusses directions for future work.

\section{AC-OPF Reformulations and Relaxations} \label{sec:MethodSOC}

The AC-OPF is a nonlinear and nonconvex optimization problem, which aims at determining the least-cost, optimal generation dispatch satisfying all demand under consideration of generator active and reactive power, line flow and nodal voltage magnitude limits \cite{Ferc_2012}. It is commonly defined in the space of $\mathbf{x} := \{\mathbf{P},\mathbf{Q},\mathbf{V},\mathbf{\theta}\}$ variables, which are defined per node and represent active power injections, reactive power injections, voltage magnitudes and voltage angles, respectively. Thus, set $\mathbf{x}$ consists of $4 \lvert \mathcal{N} \rvert$ optimization variables, where $\mathcal{N}$ denotes the set of network nodes. Bold letters indicate vectors or matrices.

Alternatively, the AC-OPF can be represented using an extended and modified set of optimization variables of size $(4 \lvert \mathcal{N} \rvert + 2 \lvert \mathcal{L} \rvert)$ \cite{EXPOSITO_RAMOS,JABR_MESHED,StrongSOCP}, where $\mathcal{L}$ denotes the set of lines (i.e., network edges). New variables are introduced to capture the nonlinearities and nonconvexities of the AC power flow equations: (a) $u_{i} := V_{i}^{2}$, (b) $c_{l} := V_{i}V_{j}\cos(\theta_{ij})$ and (c) $s_{l} := -V_{i}V_{j}\sin(\theta_{ij})$, where each transmission line $l \in \mathcal{L}$ is associated with a tuple $(i,j)$ defining its sending and receiving node. As a result, the AC-OPF is transformed from the space of $\mathbf{x} := \{\mathbf{P},\mathbf{Q},\mathbf{V},\mathbf{\theta}\}$ variables to the space of $\mathbf{y} := \{\mathbf{P},\mathbf{Q},\mathbf{u},\mathbf{\theta},\mathbf{c},\mathbf{s}\}$ variables and is given by
\begin{align}
  \min_{\mathbf{y}} \; & \sum_{i \in \mathcal{G}} c_{i}^{G} \Big( P_{i}^{G} \Big) & \label{eq:SOCOPF_OBJ} \\
  \textrm{s.t.} \; & P_{i} = G_{ii}u_{i} + \sum_{l=(i,j)} \Big( G_{ij}c_{l} - B_{ij}s_{l} \Big) & \nonumber \\
  & + \sum_{l=(j,i)} \Big( G_{ij}c_{l} + B_{ij}s_{l} \Big), & \forall i \in \mathcal{N} \label{eq:SOCOPF_PBAL} \\
  & Q_{i} = -B_{ii}u_{i} - \sum_{l=(i,j)} \Big( B_{ij}c_{l} + G_{ij}s_{l} \Big) & \nonumber \\
  & - \sum_{l=(j,i)} \Big( G_{ij}c_{l} - B_{ij}s_{l} \Big), & \forall i \in \mathcal{N} \label{eq:SOCOPF_QBAL} \\
  & 0 = c_{l}^{2} + s_{l}^{2} - u_{i}u_{j}, & \forall l \in \mathcal{L}, \label{eq:quadequal} \\
  & 0 = \theta_{j} - \theta_{i} - \text{atan} \Big( \frac{s_{l}}{c_{l}} \Big), & \forall l \in \mathcal{L}, \label{eq:ATAN2} \\
  & S_{ij}^{2} \leq (\overline{S_l})^{2}, \quad S_{ji}^{2} \leq (\overline{S_l})^{2}, & \forall l \in \mathcal{L}, \label{eq:FLOW_LIMITS} \\
  & \underline{V_i}^{2} \leq u_{i} \leq \overline{V_i}^{2}, & \forall i \in \mathcal{N}, \label{eq:C_ii_LIMITS} \\
  & \underline{P_i^{G}} \leq P_{i}^{G} \leq \overline{P_i^{G}}, \quad \underline{Q_i^{G}} \leq Q_{i}^{G} \leq \overline{Q_i^{G}}, & \forall i \in \mathcal{G}, \label{eq:PQ_LIMITS} \\
  & -\overline{V_iV_j} \leq c_{l}, s_{l} \leq \overline{V_iV_j}, & \forall l \in \mathcal{L}, \label{eq:CS_LIMITS} \\
  & \theta_{ref} = 0. &  \label{eq:ANGREF}
\end{align}

The objective function \eqref{eq:SOCOPF_OBJ} minimizes active power generation costs. Superscript $G$ denotes the contribution of conventional generators to the power injection $P_{i}$ and $Q_{i}$ at node $i$ (summarized in vectors $\mathbf{P}$ and $\mathbf{Q}$ for all nodes), while $\mathcal{G} \subseteq \mathcal{N}$ contains only nodes, which have conventional generators connected to them. Constraints \eqref{eq:SOCOPF_PBAL} and \eqref{eq:SOCOPF_QBAL} represent nodal active and reactive power balance equations, respectively. Equation \eqref{eq:quadequal} arises from the variable transformation, while voltage angles are reintroduced through constraint \eqref{eq:ATAN2}. The latter can be omitted for radial networks. Constraints \eqref{eq:C_ii_LIMITS} -- \eqref{eq:CS_LIMITS} limit the decision variables within their upper and lower bounds{\color{rev1}, which are denoted with over- and underlines}. The two inequalities in \eqref{eq:FLOW_LIMITS} constrain the apparent power flow in both directions of the line, where $S_{ij}^{2}$ (and analogously $S_{ji}^{2}$) is defined as
\begin{equation}
\begin{aligned}
S_{ij}^{2} ={} & P_{ij}^{2} + Q_{ij}^{2} \\
={}  & \Big( -G_{ij}u_{i} + G_{ij}c_{l} - B_{ij}s_{l} \Big)^{2} \\
& + \Big( (B_{ij}-B_{ij}^{sh})u_{i} - B_{ij}c_{l} - G_{ij}s_{l} \Big)^{2}, \: \forall l \in \mathcal{L}.
\end{aligned}
\end{equation}
Note that we assume a $\pi$-model of the transmission line with reactive shunt elements $B_{ij}^{sh}$ only. Equation \eqref{eq:ANGREF} sets the voltage angle of the reference bus to zero.

The optimization problem \eqref{eq:SOCOPF_OBJ} -- \eqref{eq:ANGREF} is an exact reformulation of the original AC-OPF and still nonlinear and nonconvex. However, when relaxing constraint {\color{rev1}\eqref{eq:quadequal} and approximating \eqref{eq:ATAN2}}, the original AC-OPF can be approximated by a convex quadratic optimization problem, which can be solved to global optimality. To this end, equation \eqref{eq:quadequal} is replaced by its convex second-order cone representation: $c_{ij}^{2} + s_{ij}^{2} \leq u_{i}u_{j}$, while \eqref{eq:ATAN2} can be linearized using a Taylor series expansion as proposed in \cite{JABR_MESHED} resulting in an iterative conic algorithm. The convergence is determined by the change in $\mathbf{c}$ and $\mathbf{s}$ variables, e.g., $\lvert \lvert \mathbf{c}^{\nu}-\mathbf{c}^{\nu-1} \rvert \rvert_{\infty}$
{\color{rev1}, where $\nu$ denotes the iteration counter}. Alternative convex approximations to \eqref{eq:ATAN2} have also been proposed in \cite{StrongSOCP}. {\color{rev1} Given that we reintroduce the angle constraint \eqref{eq:ATAN2}, the OPF no longer represents a pure relaxation but an approximation of the original problem.} We refer to the OPF {\color{rev1}based on relaxations and approximations} as \textit{Second-Order Cone OPF} (SOC-OPF). Note that \eqref{eq:FLOW_LIMITS} is already a convex second-order cone constraint and does not need to be reformulated. As the SOC-OPF is an {\color{rev1}approximation} of the AC-OPF, identified solutions might not be feasible to the original problem. We address this issue in Section \ref{sec:POPF}, where we propose an ex post AC feasibility recovery based on an AC power flow analysis, while in Section \ref{sec:CaseStudy} we demonstrate in our case study how the proposed procedure is not only able to recover the AC-OPF solution of a nonlinear solver but can also identify better solutions.

\section{Probabilistic Optimal Power Flow} \label{sec:POPF}
The chance-constrained OPF restricts the feasible space to a desired confidence region (CR) and identifies optimal decisions for the forecasted operating point, such that for any realization of the uncertainty and appropriate remedial actions all constraints are satisfied with a desired probability. Remedial or corrective control actions can be either pre-determined or embedded as optimization variables in the chance-constrained OPF.

\subsection{Chance-constrained SOC Optimal Power Flow}
In this paper, we propose the first formulation of a CC-SOC-OPF, which avoids the nonconvexities and convergence issues of the chance-constrained AC-OPF \cite{Line_CCACOPF} and can be computationally more efficient than other convex formulations of chance-constrained AC-OPF problems \cite{Andreas_CCACOPF}. Similar to the literature, we assume wind power generation $\mathbf{\tilde{P}^W}$ to be the only source of uncertainty. The actual wind realization $\tilde{P}_{i}^W$ is modeled as the sum of forecasted value $P_{i}^{W}$ and deviation $\xi_{i}$,
\begin{gather}
\tilde{P}_{i}^{W} = P_{i}^W + \xi_{i}, \qquad \forall i \in \mathcal{W}. \label{eq:WIND}
\end{gather}
$\mathcal{W} \subseteq \mathcal{N}$ denotes the set of nodes containing wind generators, while superscript $W$ refers to the contribution of wind power to the nodal power injection at node $i$. Recently, grid codes also require renewable energy generators to be able to provide reactive power \cite{GridCodeWindFarm}. We include the reactive power generation of wind farms as optimization variables {\color{rev1}and assume that the reactive power output follows the deviation of the active power output according to the optimal power factor $\cos \phi$ at the forecasted operating point. Thus, the actual realization of the reactive wind power output is modeled as follows
\begin{gather}
\tilde{Q}_{i}^{W} = \lambda (P_{i}^{W} + \xi_i), \qquad \forall i \in \mathcal{W},
\end{gather}
where $\lambda := \sqrt{\frac{1-\cos{\phi}^{2}}{\cos{\phi}^{2}}}$ is an optimization variable and denotes the ratio between reactive and active wind power generation.}

We model all decision variables $\mathbf{\tilde{y}(\xi)}$ of the OPF as functions of the uncertainty $\mathbf{\xi}$: $\mathbf{\tilde{y}(\xi)} = \mathbf{y} + \mathbf{\Delta y(\xi)}$, where $\mathbf{y}$ represents the optimal setpoint at the forecasted operating point and $\mathbf{\Delta y(\xi)}$ the system response to a change in active power injection (i.e., wind power deviation $\xi$). The chance-constrained OPF minimizes the total generation cost for the forecasted operating point and is formulated as follows
{\color{rev2}
\begin{align}
\min_{\mathbf{y}} \; & \sum_{i \in \mathcal{G}} c_{i}^{G} \Big( P_{i}^{G} \Big)& \label{eq:CCSOCOPF_OBJ}   \\
\textrm{s.t.} \; & \textrm{\eqref{eq:SOCOPF_PBAL} -- \eqref{eq:ATAN2}, \eqref{eq:ANGREF}} & \textrm{for } \mathbf{y},  \label{eq:CC_EQ} \\
& \mathbb{P} \Big(  \tilde{y}_{i}(\xi) \leq \overline{y_i} \Big) \geq 1- \epsilon \quad & \forall \tilde{y}_i(\xi) \in \mathbf{\tilde{y}(\xi)}, \label{eq:CC_INEQ1}  \\
& \mathbb{P} \Big(  \tilde{y}_{i}(\xi) \geq \underline{y_i} \Big) \geq 1- \epsilon \quad & \forall \tilde{y}_i(\xi) \in \mathbf{\tilde{y}(\xi)}, \label{eq:CC_INEQ2} \\
& \mathbb{P} \Big(  \tilde{S}_{ij}^{2}(\xi) \leq (\overline{S_l})^{2} \Big) \geq 1- \epsilon \quad & \forall l \in \mathcal{L}, \label{eq:CC_INEQ3} \\
& \mathbb{P} \Big(  \tilde{S}_{ji}^{2}(\xi) \leq (\overline{S_l})^{2} \Big) \geq 1- \epsilon \quad & \forall l \in \mathcal{L},  \label{eq:CC_INEQ4}
\end{align}
}
where $\epsilon \in (0,1)$ represents the allowed constraint violation probability. Thus, the CR (i.e., the restricted feasible space) of the chance-constrained OPF is defined by the confidence level $(1-\epsilon)$.

{\color{rev1}Note that \eqref{eq:CC_INEQ1} -- \eqref{eq:CC_INEQ4} represent separate chance constraints, i.e., the probability of satisfying \eqref{eq:FLOW_LIMITS} -- \eqref{eq:CS_LIMITS} is enforced for each constraint individually and not jointly. We use separate chance constraints as they (i) do not significantly change the computational complexity of the problem as opposed to joint formulations and (ii) have proven to also effectively reduce the joint violation probability, while remaining less conservative than approaches which explicitly target joint chance constraints and usually overly satisfy them \cite{Line_PhD,Andreas_CCACOPF}. Separate chance constraints are also used to approximate joint chance constraints \cite{Nemirovski_Shapiro}. They offer the flexibility to identify and target individual constraints which are decisive for the system's security, while avoiding to unnecessarily limit the solution space along other dimensions that are of minor significance to security but could have a substantial impact on costs.}

Problem \eqref{eq:CCSOCOPF_OBJ} -- \eqref{eq:CC_INEQ4} represents the chance-constrained formulation of the exact AC-OPF \eqref{eq:SOCOPF_OBJ} -- \eqref{eq:ANGREF} and can be relaxed as described in Section \ref{sec:MethodSOC} to obtain the convex CC-SOC-OPF. All equality constraints (i.e., \eqref{eq:SOCOPF_PBAL} -- \eqref{eq:ATAN2}, \eqref{eq:ANGREF}) and relaxations of them are considered for the forecasted operating point only, as including \eqref{eq:WIND} and $\mathbf{\tilde{y}(\xi)}$ directly in \eqref{eq:CC_EQ} would render the problem semi-infinite and thus, intractable.
\subsection{Control Policies: Modeling the System Response}
In order to approximately model the system response to a change in wind power injection $\xi$, we use linear policies for all variables concerned.
\subsubsection{Reserve Deployment}
Fluctuations in active power generation are balanced by conventional generators, which are assumed to provide up- and down-reserves according to their generator participation factors $\mathbf{\gamma}$. The participation factors are pre-determined and proportional to each generator's installed capacity with respect to the total installed capacity {\color{rev1}of conventional generation}. The generator output is adjusted according to the total power mismatch $\Xi = \sum_{i \in \mathcal{W}} \xi_{i}$ \cite{Line_AnalyticalReformulation}. Hence, the sum of all generator contributions to the reserve deployment needs to balance the total power mismatch $\Xi$, which implies the following condition: $\sum_{i \in \mathcal{G}} \gamma_{i} = 1$, so that the total contribution of all generators equals the total power mismatch $\sum_{i \in \mathcal{G}} \gamma_{i} \sum_{i \in \mathcal{W}} \xi_{i} = \Xi$. {\color{rev1}Similar to \eqref{eq:WIND}, the actual dispatch of a conventional unit is modeled as the sum of its optimal dispatch at the expected wind infeed and its reaction to the wind power deviation,
\begin{align}
\tilde{P}_{i}^{G}(\mathbf{\xi}) & = P_{i}^G + \Delta P_{i}^{G}(\mathbf{\xi}) & \nonumber \\
& = P_{i}^G - \gamma_{i} \Xi + \Delta P_{i}^{U} (\mathbf{\xi}), & \forall i \in \mathcal{G}. \label{eq:RESERVES}
\end{align}
$\Delta P_{i}^{U}(\mathbf{\xi})$ represents the unknown nonlinear changes in active power losses, which are usually compensated by the generator at the reference bus. Thus, this term is equal to zero for all other generators. As for the other variables $\mathbf{Q}$, $\mathbf{u}$, $\mathbf{c}$, $\mathbf{s}$, $\mathbf{\theta}$, which vary nonlinearly with the wind power injection, we approximate $\Delta P_{i}^{U}$ through a linearization around the forecasted operating point, which is described in the next section.} Note that the participation factors can also be included as optimization variables and defined for each wind infeed individually. However, a higher number of optimization variables and additional second-order cone constraints in that case also increase the computational burden.

\subsubsection{Linear Decision Rules} \label{sec:LDR}
We derive linear sensitivities of each variable with respect to the uncertainty based on a Taylor series expansion around the forecasted operating point. We model the response as follows: $\mathbf{\Delta y(\xi) = \frac{\partial y}{\partial \xi} \xi =  \Upsilon \xi}$, such that $\mathbf{\tilde{y}(\xi) = y + \Delta y(\xi)}$ represents a linear decision rule (LDR) with respect to the uncertainty. The authors in \cite{Line_CCACOPF,Jeremias_CCACOPF,Hao_UMs} have derived the linear sensitivity factors from the Jacobian matrix at the forecasted operating point of the original AC power flow equations. The detailed derivation can be found in \cite{Line_PhD}. In this work, we derive the linear sensitivity factors $\mathbf{\Upsilon}$ based on the Jacobian matrix of the alternative load flow equations \eqref{eq:SOCOPF_PBAL} -- \eqref{eq:ATAN2}, such that we can directly use them as input to the convex chance-contrained SOC-OPF,
\begin{gather}
\begin{bmatrix}
\mathbf{\Delta P} \\
\mathbf{\Delta Q} \\
\mathbf{0} \\
\mathbf{0}
\end{bmatrix} =
\begin{bmatrix}
\mathbf{J^{SOC}}
\end{bmatrix} \Bigg\rvert_{\mathbf{y}}
\begin{bmatrix}
\mathbf{\Delta u} \\
\mathbf{\Delta c} \\
\mathbf{\Delta s} \\
\mathbf{\Delta \theta}
\end{bmatrix}. \label{eq:JACSOC}
\end{gather}

{\color{rev1}The derivation of $\mathbf{\Upsilon}$ is presented in the appendix.} The left-hand side of equation \eqref{eq:JACSOC} can also be expressed in terms of the uncertain wind infeed, the generator participation factors, the optimal ratio between reactive and active wind power generation at the forecasted operating point and the unknown nonlinear changes in active and reactive power. We replace the entries for $\mathbf{\Delta P}$ and $\mathbf{\Delta Q}$ accordingly and {\color{rev1}modify} the system of equations considering the following assumptions aligned with current practices in power system operations:
\begin{itemize}
  \item the change in active power losses is compensated by the generator at the reference bus: {\color{rev1}$\mathbf{ \Delta P_{PV,PQ}^{U} = 0}$};
  \item changes in reactive power generation are compensated by generators at PV and reference buses, as PQ buses are assumed to keep their active and reactive power injection constant: $\mathbf{ \Delta Q_{PQ} = 0}$;
  \item generators at PV and reference buses regulate their reactive power output to keep the voltage magnitude and thus, the square of the voltage magnitude constant: $\mathbf{ \Delta u_{PV,ref} = 0}$;
  \item the voltage angle at the reference bus is always zero: $\Delta \theta_{ref} = 0$.
\end{itemize}

Rearranging the resulting system of equations allows us to define the changes in all variables of interest (i.e., $\mathbf{\Delta y} \setminus \{ \mathbf{\Delta P_{PV}^{U}}, \mathbf{\Delta P_{PQ}^{U}}, \mathbf{\Delta Q_{PQ}}, \mathbf{\Delta u_{PV}}, \Delta u_{ref}, \Delta \theta_{ref} \}$) as a function of $\mathbf{\xi}$.
The changes in active and reactive branch flows due to fluctuations in wind infeed can be represented by a linear combination of the changes in $u$, $c$ and $s$ variables, as shown in Eq. \eqref{eq:DeltaPij} for active branch flows.
\begin{align}
\Delta P_{ij} (\xi) ={} & -G_{ij} \Delta u_{i} + G_{ij} \Delta c_{l} - B_{ij} \Delta s_{l}, & \forall l \in \mathcal{L}. \label{eq:DeltaPij}
\end{align}
The chance constraints of the apparent branch flow constraints are thus formulated as quadratic chance constraints for all lines $l:=(i,j)$ (and analogously for the reversed power flow direction $(j,i)$),
\begin{multline}
\mathbb{P} \Big[ \Big( P_{ij} + \Delta P_{ij} (\xi) \Big)^{2} \\
+ \Big( Q_{ij} + \Delta Q_{ij} (\xi) \Big)^{2} \leq \Big(\overline{S_{l}}\Big)^2 \Big] \geq 1-\epsilon. \label{eq:QUAD_CC}
\end{multline}
\subsubsection{Reformulating the Linear Chance Constraints} \label{sec:LinCC}
The LDR approach coupled with the assumption that the wind deviations $\mathbf{\xi}$ follow a multivariate distribution with {\color{rev1}known mean and covariance} allows us to analytically reformulate the single chance constraints in \eqref{eq:CC_INEQ1} -- \eqref{eq:CC_INEQ4} to deterministic constraints \cite{Line_AnalyticalReformulation}. We choose the analytical approach based on a Gaussian distribution {\color{rev1}with zero mean} given that previous work in \cite{Line_CCACOPF} has shown that (i) it is reasonably accurate, even when the uncertainty is not normally distributed, and (ii) it performs better than sample-based reformulations based on Monte Carlo simulations and the so-called scenario approach \cite{Kostas_OnTheRoad}. Using the properties of the Gaussian distribution, the linear chance constraint {\color{rev1}$\mathbb{P}[ y_i + \mathbf{\Upsilon_{i} \xi} \leq \overline{y_{i}} ] \geq 1 - \epsilon$} is reformulated as follows:
{\color{rev1}
\begin{align}
y_i + \Phi^{-1}(1-\epsilon) \sqrt{\mathbf{\Upsilon_{i} \Sigma \Upsilon_{i}^{T}}} \leq \overline{y_{i}}, \label{eq:LIN_CC}
\end{align}}
{\color{rev1}where $\Phi^{-1}$ denotes the inverse cumulative distribution function of the Gaussian distribution and $\mathbf{\Sigma}$ the ($\rvert \mathcal{W} \lvert \times \rvert \mathcal{W} \lvert$) covariance matrix. Note that {\color{rev1}$\mathbf{\Upsilon_{i}}$ denotes the i-th row of matrix $\mathbf{\Upsilon}$} and is a ($1 \times \rvert \mathcal{W} \lvert$) vector containing the sensitivity of the considered variable w.r.t. $\xi$ at each node in $\mathcal{W}$. The derivation of how the chance constraint is reformulated to its deterministic form can be found in e.g., \cite{Line_PhD}.} It can be observed that introducing uncertainties results in a tightening of the original constraint {\color{rev1}$y_{i} \leq \overline{y_{i}}$} and thus, a reduction of the feasible space to the CR defined by the confidence level $(1-\epsilon)$. The introduced margin {\color{rev1}$\Omega_{i} = \Phi^{-1}(1-\epsilon) \sqrt{\mathbf{\Upsilon_{i} \Sigma \Upsilon_{i}^T}}$} secures the system against uncertain infeeds and was termed \textit{uncertainty margin} in \cite{Line_AnalyticalReformulation}.

\subsubsection{Reformulating the Quadratic Chance Constraints} \label{sec:Quad_CC}
The apparent flow constraint inside \eqref{eq:QUAD_CC} is indeed convex but nonlinear, which prevents a straight-forward analytical reformulation of the chance constraint similar to the linear one in \eqref{eq:LIN_CC}. We therefore approximate the quadratic chance constraint by a set of probabilistic absolute value constraints and a nonprobabilistic quadratic constraint as proposed in \cite{TwoSided_LinCC} and recently applied in \cite{miles_yuri_line}. Constraint \eqref{eq:QUAD_CC} is replaced by the following set of constraints:
\begin{align}
\mathbb{P} \Big[ \lvert P_{ij} + \Delta P_{ij}(\xi) \rvert \leq k_{ij}^{P} \Big] & \geq1- \beta \epsilon, \label{eq:ABS1} \\
\mathbb{P} \Big[ \lvert Q_{ij} + \Delta Q_{ij}(\xi) \rvert \leq k_{ij}^{Q} \Big] & \geq1- (1-\beta) \epsilon, \label{eq:ABS2} \\
(k_{ij}^{P})^2 + (k_{ij}^{Q})^2 & \leq (\overline{S_l})^2. \label{eq:fP2fQ2}
\end{align}
$k_{ij}^{P}$ and $k_{ij}^{Q}$ are optimization variables introduced to enable the reformulation. The absolute value constraints \eqref{eq:ABS1} and \eqref{eq:ABS2}, also called two-sided linear chance constraints, are a special type of joint chance constraints and can be approximated by two single linear chance constraints, e.g. $\mathbb{P} [ P_{ij} + \Delta P_{ij}(\xi) \leq k_{ij}^{P} ] \geq 1-\beta\epsilon$ and $\mathbb{P} [ P_{ij} + \Delta P_{ij}(\xi) \geq -k_{ij}^{P}] \geq 1-\beta\epsilon$. In this form, the constraints can be reformulated analytically as desribed in Section \ref{sec:LinCC}. $\beta \in (0,1)$ is a parameter, which balances the trade-off between violations in the two constraints \eqref{eq:ABS1} and \eqref{eq:ABS2} and ensures that the union of the constraints still satisfies the desired confidence level, i.e., $\mathbb{P} [ \eqref{eq:ABS1} \cup \eqref{eq:ABS2} ] \geq 1-\epsilon$. Note that without $\beta$, i.e., when enforcing \eqref{eq:ABS1} and \eqref{eq:ABS2} with $(1-\epsilon)$, respectively, the union $\mathbb{P} [ \eqref{eq:ABS1} \cup \eqref{eq:ABS2} ]$ only holds with $(1-2\epsilon)$ \cite{TwoSided_LinCC}.

{\color{rev1}The major benefit of using an approach combining LDRs and analytical reformulations lies in its adaptability to a wide range of uncertainty behavior. The authors in \cite{LineDan_IREP,Line_DistributionallyRobustSCOPF} thoroughly discuss how different assumptions on the statistical behavior of the uncertainty can be incorporated into the analytical reformulation. To this end, \eqref{eq:LIN_CC} can be generalized by replacing $\Phi^{-1}(1-\epsilon)$ with a more general function $f_{\mathcal{P}}^{-1}(1-\epsilon)$, whose value can be determined for any distribution if the mean $\mu$ and variance $\Sigma$ of the uncertainty are known. The exact expressions of $f_{\mathcal{P}}^{-1}(1-\epsilon)$ for different distributions are derived in \cite{Line_DistributionallyRobustSCOPF}.
Different values for $f_{\mathcal{P}}^{-1}(1-\epsilon)$ and thus, different assumptions on the distribution are simply reflected in the optimization through different values for the uncertainty margins $\Omega_i = f_{\mathcal{P}}^{-1}(1-\epsilon)\sqrt{\mathbf{\Upsilon_{i} \Sigma \Upsilon_{i}^{T}}}$. 
}

\subsubsection{Modeling Inaccuracies}
\begin{figure}
\centering
\includegraphics[scale=0.2]{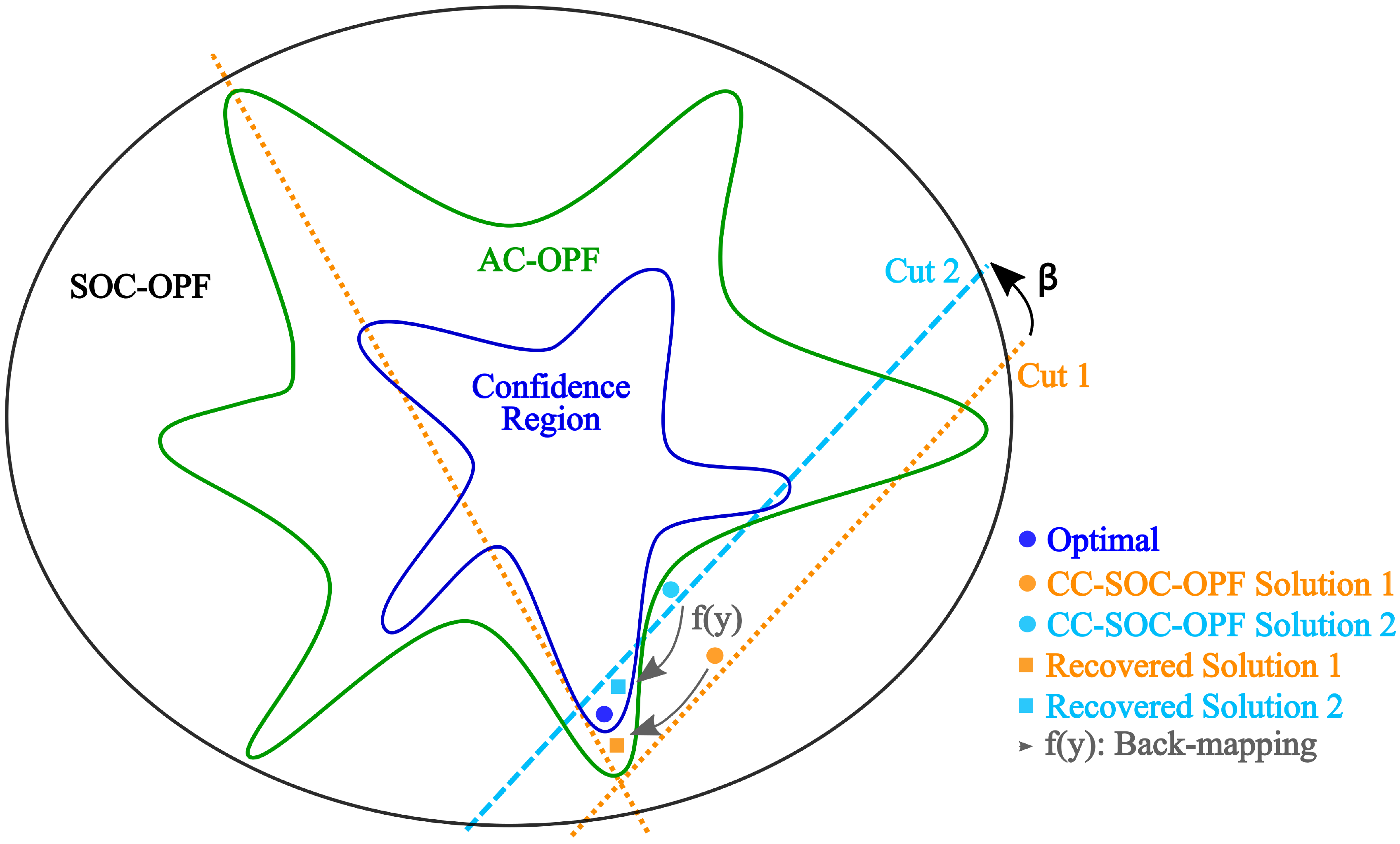}
\centering \caption{Illustration of how modeling inaccuracies might affect the CC-SOC-OPF and visualization of the AC feasibility recovery (back-mapping).}
\label{fig:FeasibleSpaceCCSOC}
\end{figure}

The linearization of the uncertainty impact and the approximation of {\color{rev1}both} the quadratic chance constraints {\color{rev1}and the angle constraint} are sources of inaccuracies and entail that the CC-SOC-OPF solution might still lie outside the feasible space of the AC-OPF and the CR (i.e., the chance-constrained AC-OPF) despite the constraint tightenings as depicted in Fig. \ref{fig:FeasibleSpaceCCSOC}. This highlights the need for appropriate back-mapping procedures to project the CC-SOC-OPF solution back into the feasible space through either ex ante relaxation tightenings or an ex post power flow analysis. Tightenings improve the relaxation but can still not guarantee AC feasibility of the solution. Therefore, we propose to use the solution of the relaxed OPF as a warm start to an AC power flow analysis. To ensure that the CC-SOC-OPF solution is not only projected back into the AC feasible space but into the CR, increased levels of conservatism are required in the CC-SOC-OPF modeling, where $\beta$ provides an additional degree of freedom to tighten the relaxation along the dimension of the corresponding quadratic chance constraint. Note that the CC-SOC-OPF solution might still not be AC feasible due to loose bounds along other dimensions, but can be made so through the feasibility recovery. How to appropriately choose $\beta$ has to our knowledge not been addressed in previous work. Performing a rigorous investigation in our case studies, we find that for $\mathbb{P} [ \eqref{eq:ABS1} \cup \eqref{eq:ABS2} ] \geq 1-\epsilon$ to hold while keeping the additional cost incurred by the uncertainty as low as possible, $\beta$ needs to be tuned for each quadratic chance constraint individually. {\color{rev1}Alternatively, choosing a value for $\beta$ of 0.5, as done in \cite{miles_yuri_line}, provides a convex inner approximation of the quadratic chance constraint and thus, a robust approximation of the constraint \cite{TwoSided_LinCC}. This is aligned with the classical Bonferroni approximation, which uses the union bound to approximate the violation probability $\epsilon$ of $K$ jointly considered chance constraints by $K$ single chance constraints, each of which is enforced by $(1-\frac{\epsilon}{K})$ \cite{Nemirovski_Shapiro}.}

\subsubsection{{\color{rev1}Critical Line} Screening}
In order to reduce both the effort associated with the parameter tuning and the number of new variables and constraints, which need to be introduced to reformulate \eqref{eq:QUAD_CC}, we propose to perform a pre-screening based on the forecasted operating point {\color{rev1}to identify the most critical lines}. Specifically, we evaluate the vertices $\mathbf{\kappa}$ of the polyhedral outer approximation of the ellipsoidal uncertainty set given by the multivariate Gaussian distribution \cite{Andreas_CCACOPF}, as one of the vertices includes the worst-case realization of the ellipsoidal uncertainty set. We then use a linearization based on Power Transfer Distribution Factors (PTDF) to approximate the change in active power line flows at each vertex, i.e., $\mathbf{\Delta PF}^{\kappa} = \mathbf{PTDF \times \Delta P}^{\kappa}$, where the change in active power injection $\mathbf{\Delta P}^{\kappa}$ is defined w.r.t. the forecasted operating point. The final active power flows at each vertex show which lines could be overloaded and thus, have a high risk of exceeding the allowable violation probability. These lines are classified as \textit{critical} and their capacity constraints are included as chance constraints. The branch flows on all other lines are constrained by their usual limits and do not consider an uncertainty margin. This procedure takes place iteratively after every solution of a CC-SOC-OPF until no new critical lines are identified.

\subsection{Solution Algorithm}
The sensitivity factors $\mathbf{\Upsilon}$ depend nonlinearly on the operating point and would render the problem nonconvex if they were introduced as optimization variables. Therefore, we define $\mathbf{\Upsilon}$ and the uncertainty margins $\mathbf{\Omega}$ outside the optimization problem and adopt the iterative solution algorithm from \cite{Jeremias_CCACOPF} and apply it in the context of a SOC-OPF, which allows us to maintain the convexity of the CC-SOC-OPF. We improve on the work in \cite{Jeremias_CCACOPF} and \cite{Line_CCACOPF} by avoiding nonconvexities in the optimization and thus, provide convergence guarantees for the iterative solution algorithm. The algorithm converges as soon as the change in $\mathbf{\Omega}$ between two consecutive iterations is lower than a pre-defined tolerance value $\rho$ and is defined as follows:
\begin{algorithmic}[1]
\STATE Set iteration count: $\nu \leftarrow 0$
\WHILE{$\lvert \lvert \mathbf{\Omega}^{\nu}-\mathbf{\Omega}^{\nu-1} \rvert \rvert_{\infty} > \rho$}
\IF{$\nu=0$}
\STATE solve the SOC-OPF for the forecasted wind infeed without considering uncertainty and obtain the operating point $\mathbf{y}^{0}$
\STATE evaluate $\mathbf{\Upsilon^{0}}$ and $\mathbf{\Omega^{0}}$ at $\mathbf{y}^{0}$
\ENDIF
\STATE perform {\color{rev1}critical line} screening based on $\mathbf{y^{\nu}}$ and $\kappa$ and append the \textit{critical} line list
\STATE include $\mathbf{\Omega^{\nu}}$ according to \eqref{eq:LIN_CC} for all variables $\mathbf{y(\xi)}$ and \eqref{eq:ABS1}-\eqref{eq:fP2fQ2} for all \textit{critical} lines
\STATE solve CC-SOC-OPF to obtain $\mathbf{y}^{\nu+1}$
\STATE evaluate $\mathbf{\Upsilon^{\nu+1}}$ and $\mathbf{\Omega^{\nu+1}}$ at $\mathbf{y^{\nu+1}}$
\STATE $\nu \leftarrow \nu+1$
\ENDWHILE.
\end{algorithmic}
This allows us to fully exploit the efficiency of solvers for convex programming. Note that the iterative solution algorithm for the chance constraints adds an additional outer iteration loop to the iterative conic procedure for approximating the angle constraint \eqref{eq:ATAN2}.

{\color{rev1}
\subsection{Robustness and Extensions of the Algorithm} \label{sec:robustness_extensions}

In this Section, we discuss several aspects and possible extensions of the algorithm which are not only limited to the examples mentioned here. Other possible extensions include the consideration of security requirements (e.g., N-1 and stability criteria) and distributionally robust formulations (see Section~\ref{sec:FutureWork}).

\subsubsection{The Need for Robustness}
The efficiency of the iterative algorithm to handle large dimensions of the uncertainty has been demonstrated in \cite{Line_CCACOPF}, where the full AC power flow equations and the Polish test case of 2383 buses with 941 uncertain loads are used. The example also proves the good performance of the iterative approach even for nonconvex problems, if it converges. This is emphasized by the work in \cite{LineDan_IREP}, where the authors analyze the impact of perturbations of the initial operating point on the final solution. Despite having large differences in cost and uncertainty margins in the first iteration, the results quickly converge to solutions, which share the same cost and uncertainty margins. Nevertheless, several instances were also identified in \cite{LineDan_IREP}, where the iterative algorithm failed to converge as a result of the nonconvexities. Some cases encountered infeasibility of the OPF at intermediate iterations and failed to recover subsequently. Others exhibited a cycling behavior (e.g., the five bus case from \cite{Waqqash_LocalSolutionsACOPF}), where the algorithm oscillated between two different local optima, which had large differences in their corresponding uncertainty margins and were located in two disjoint regions of the feasible space.

The recent work in \cite{Anders_RobustnessAndScalabilityOfSDP} compares the performance of several convex solvers with three nonlinear solvers by solving nonprobabilistic AC-OPF relaxations based on SDP and standard AC-OPFs for 133 different test cases of up to 25'000 buses, respectively. Contrary to the SDP solvers, even the most efficient nonlinear AC-OPF solver failed to converge to a solution for 20 out of the 133 systems tested including all test cases over 10'000 buses. All the examples mentioned highlight the need for more robust solution approaches and the ability of convex programming to provide them. The results from \cite{Anders_RobustnessAndScalabilityOfSDP} are proof that robustness is not only an issue for more sophisticated AC-OPF algorithms, which include functionalities beyond the usual ones from the standard AC-OPF. Robustness is already a challenge for the standard AC-OPF, whose convergence and success is fundamental to the functioning of any other algorithm that builds on top of it.

\subsubsection{Uncertainty Dimension}
Besides maintaining convexity, another benefit of combining the iterative approach, where $\mathbf{\Upsilon}$ is computed outside the OPF and the analytical reformulation of the chance constraints is the independence from the size of the uncertainty set $\lvert \mathcal{W} \rvert$. $\lvert \mathcal{W} \rvert$ solely has an impact on the dimensions of $\mathbf{\Upsilon}$, $\mathbf{\Sigma}$ and $\mathbf{\xi}$. Operations on these matrices are only conducted at steps 5 and 10 of each iteration, which are decoupled from the optimization in steps 4 and 9. The computational complexity of the approach is mainly determined by the size of the optimization problem, which remains unchanged with an increasing number of uncertainty sources. It only changes if more \textit{critical lines} are detected during the critical line screening, whose reformulated chance constraints need to be added to the constraint set. However, given that the number of active constraints is usually low, even in large systems, this is not expected to be an obstacle \cite{Bienstock_RiskAwareNetworkCtrl}.}

{\color{rev1}
\subsubsection{Large Uncertainty Ranges}
The linearization of the uncertainty impact performs better close to the forecasted operating point and might lead to inaccuracies, when considering large ranges of the uncertainty. Given that we operate in a nonlinear space, some type of affine approximation is necessary to keep the chance constraints tractable. However, we do not expect that this poses a significant limitation for this method. Our approach is expected to be used for power system operations, which usually require short-term forecasts (e.g. usually days/hours instead of months or years). The forecast uncertainty range associated with such time intervals is expected to be reasonable for our method.}

{\color{rev1}
\subsubsection{Integer Variables}
Given that the resulting optimization problem solved at each iteration is formulated in almost the same way as the deterministic SOC-OPF with the only difference of having tighter variable limits to represent the uncertainty impact, the problem can also be extended to include integer variables accounting for e.g., shunt elements or tap changers. Once the optimal integer decisions at one iteration have been determined, the uncertainty margins can still be derived as described above by a linearization around the optimal integer solution. However, different integer solutions throughout the iteration process could also result in constantly changing values for $\mathbf{\Omega}$ leading to convergence issues. Similar to what has been proposed in \cite{LineDan_IREP}, one possible solution approach could be a branching algorithm, where the initial uncertainty margins of each integer solution are used to obtain a new integer solution (i.e., a new branch). At the same time, the CC-SOC-OPF algorithm described above is applied to each integer solution with the integer variables fixed to their optimal values in order to obtain the final uncertainty margins of the corresponding branch. The associated operating point is added to the list of candidate solutions. The branching algorithm would be considered to converge as soon as it does not identify any new integer solutions, which have not been explored yet. The most cost-efficient candidate solution would constitute the final solution.
}

\section{Case Study} \label{sec:CaseStudy}
We evaluate the performance of the proposed CC-SOC-OPF on the IEEE 118 bus test system \cite{TechReport_case118}. We assume the MW line ratings given in \cite{TechReport_case118} as MVA line ratings and reduce them by 30\% to obtain a more constrained system. We add wind farms to node 5 and 64 with expected production levels of 300 MW and 600 MW, respectively. We assume a standard deviation of 10\% and a power factor between 0.95 capacitive and 0.95 inductive for each wind farm. Minimum and maximum voltage limits are set to 0.94 p.u. and 1.06 p.u.. Generator cost functions are assumed to be linear.

We first demonstrate how an SOC-OPF coupled with an AC feasibility recovery is able to approach the solution of a nonlinear solver for the exact AC-OPF problem. Afterwards, we show how the convex CC-SOC-OPF coupled with the AC feasibility recovery is able to identify even better solutions in terms of operation cost than the CC-AC-OPF from \cite{Line_CCACOPF}. We evaluate the constraint violation probabilities in all cases empirically using Monte Carlo simulations of AC power flow calculations based on 10'000 scenarios drawn from a multivariate Gaussian distribution. All simulations related to SOC-OPFs were carried out in Python using the Gurobi Optimizer. The nonconvex CC-AC-OPF was implemented in Matlab, where the OPF at each iteration was solved using Matpower and its internal MIPS solver \cite{MATPWR}. All AC power flow analyses, i.e., the AC feasibility recovery and the Monte Carlo simulations, were also carried out with Matpower.
\subsection{Recovering the SOC-OPF Solution} \label{sec:AcFeasibilityRecovery}
We evaluate the SOC-OPF at the forecasted operating point without considering wind power uncertainty and compare the outcome to the standard AC-OPF solution. We assume a convergence tolerance of $10^{-6}$ for the sequential conic procedure to approximate the angle constraint \eqref{eq:ATAN2}. The objective function value of the relaxed problem is identical to the one obtained with the exact problem (37'692.03\euro), providing a seemingly tight relaxation with zero relaxation gap. However, when evaluating the full AC power flow equations at the operating point identified by the SOC-OPF, we observe a mismatch of active and reactive power injections at nodes 37 and 38, which despite being fairly small (i.e., 0.06 MW and 0.98 Mvar) indicate that the operating point is not AC feasible. The infeasibility is also reflected in the SOC constraint of line 50 connecting nodes 37 and 38, which is the only one that fails to maintain the equality constraint \eqref{eq:quadequal} at the SOC-OPF solution. This also highlights the inadequacy of defining the relaxation gap of OPF relaxations solely based on differences in objective function values. The OPF objective function only considers costs on active power generation and thus, neglects the fact that one $\mathbf{P}$ solution might be associated with numerous $\{\mathbf{Q,V,\theta}\}$ solutions, not all of which might be feasible.

Therefore, we propose to use the SOC-OPF solution as a warm start to an AC power flow analysis in order to recover the feasible AC power flow solution. However, given that power flow calculations do not consider any variable limits, we need to enforce generator reactive power limits, slack bus active power limits (by e.g. changing the slack bus if necessary), and check for voltage and branch flow limits. The final power flow solution results in a dispatch with slightly lower generation cost (37'691.97\euro).

The results of the Monte Carlo simulations are listed in Table \ref{tab:1} showing the maximum violation probabilities for generator active power, bus voltage, and apparent branch flow limits.
Table \ref{tab:1} also shows the joint violation probability, which represents the probability of at least one constraint being violated (i.e., the number of samples with at least one constraint violation out of the 10'000 tested). A joint violation probability of 100\% for the standard AC-OPF and SOC-OPF indicates that neither OPF algorithm results in an operating point, which is able to maintain feasibility for any other wind power realization if uncertainty in wind power infeed is not explicitly accounted for. The maximum violation probability of single constraints in that case lies around 50\%.
\subsection{{\color{rev1}Critical Line} Screening}
The {\color{rev1}critical line} screening shows that line 100, which is already congested at $\mathbf{y}^{0}$, violates its branch flow limits in both directions of the line.
Furthermore, the limits of line 37, which is not congested at $\mathbf{y}^{0}$, are also estimated to be violated in the positive flow direction (i.e., from node 8 to node 30). Thus, we include three quadratic chance constraints: two for line 100 defining both flow directions and one for line 37 defining only the positive flow direction. The weighting factors are denoted with $\beta^{\leftrightarrow}_{100}$ and $\beta^{\rightarrow}_{37}$ with the arrows indicating the direction of the constrained flow.
\begin{table}
  \caption{Results of the Monte Carlo simulations: Comparison of maximum violation probabilities between AC-OPF, SOC-OPF*, CC-SOC-OPF* and CC-AC-OPF. The results of the CC-SOC-OPF* include the ones obtained without $\beta$ and with the final optimal values for $\beta$.}
  \label{tab:1}
  \centering
  \begin{tabular}{ccccc}
  \toprule
  \textbf{AC-OPF} & \textbf{SOC-OPF*} & \textbf{CC-AC-OPF} & \multicolumn{2}{c}{\textbf{CC-SOC-OPF*}} \\
  & & & $\mathbf{\beta = \emptyset}$ & $\mathbf{\beta \neq \emptyset}$ \\ \midrule
    \multicolumn{5}{c}{\textbf{Generator active power limits}}\\
    48.74\% & 48.74\% & 5.00\% & 4.96\% & 4.96\% \\ \hline
    \multicolumn{5}{c}{\textbf{Bus voltage limits}}\\
    43.86\% & 56.89\% & 3.26\% & 1.30\% & 0.25\% \\ \hline
    \multicolumn{5}{c}{\textbf{Apparent power line flow limits}}\\
    50.22\% & 50.12\% & 4.07\% & 7.72\% & 5.00\% \\ \hline
    \multicolumn{5}{c}{\textbf{Joint violation probability}}\\
    100\% & 100\% & 14.85\% & 17.35\% & 15.60\%\\
  \bottomrule
  \end{tabular}
  \\
  \vspace{0.5ex}
  \centering \scriptsize * The Monte Carlo simulations were carried out with the recovered SOC solution.
\end{table}
\subsection{Chance-constrained SOC-OPF}
We first determine the optimal parameters $\beta$ for the CC-SOC-OPF and then compare it to the CC-AC-OPF algorithm proposed in \cite{Line_CCACOPF}. We assume an acceptable violation probability $\epsilon$ of 5\% for all chance constraints. The convergence tolerance $\rho$ of the uncertainty margins for the iterative CC-SOC-OPF and CC-AC-OPF is set to $10^{-5}$. Both algorithms converge after 4 iterations demonstrating the suitability of the iterative solution algorithm for both OPFs. However, the algorithm is more robust in case of the SOC-OPF due to the convexity of the problem solved at each iteration step, which provides convergence guarantees \cite{DecompositionBook1}. {\color{rev1}}
\subsubsection{Quadratic chance constraints without weighting factors $\beta=\emptyset$}
First, we analyze the solution without considering the weighting factors $\beta^{\leftrightarrow}_{100}$ and $\beta^{\rightarrow}_{37}$, i.e., we enforce the two separate absolute value constraints \eqref{eq:ABS1} -- \eqref{eq:ABS2} for each quadratic chance constraint with the usual confidence level $(1-\epsilon)$. The results are listed in Table \ref{tab:1}. It can be observed that the chance constraints for active power generation and voltage magnitudes are satisfied. However, the maximum violation probability of the apparent branch flow limits exceeds the allowable threshold of 5\% and indicate that the recovered solution is not located within the CR. Specifically, this violation is caused by the flow in the positive flow direction on line 37 and confirms that the union of the two separate constraints \eqref{eq:ABS1} and \eqref{eq:ABS2} only holds with $1-2\epsilon$ as described in Section \ref{sec:Quad_CC}. In case of line 100, the level of conservatism for enforcing the two separate constraints is already sufficient, such that the violation probabilities are reduced to 4.47\% and 4.17\% in the positive and negative flow directions, respectively.
\subsubsection{Quadratic chance constraints with weighting factors $\beta = \{\beta^{\rightarrow}_{37},\beta^{\leftrightarrow}_{100}\}$}
In order to evaluate the impact of different values of the weighting factors on the CC-SOC-OPF, we perform a sensitivity analysis varying $\beta^{\leftrightarrow}_{100}$ uniformly in both flow directions from 0.1 to 0.9 in 0.1 increments and add 0.01 and 0.99 as the approximate endpoints of the interval $\beta \in (0,1)$ (i.e., in total 11 samples). As line 37 has proven to be more critical, we use a finer sampling of $\beta^{\rightarrow}_{37}$ between 0.02 and 0.98 in 0.02 increments (i.e., 49 samples). Hence, we perform the sensitivity analysis based on 539 simulations of the CC-SOC-OPF with a subsequent AC feasibility recovery.

\begin{figure}
\centering
\includegraphics[width=3.5in]{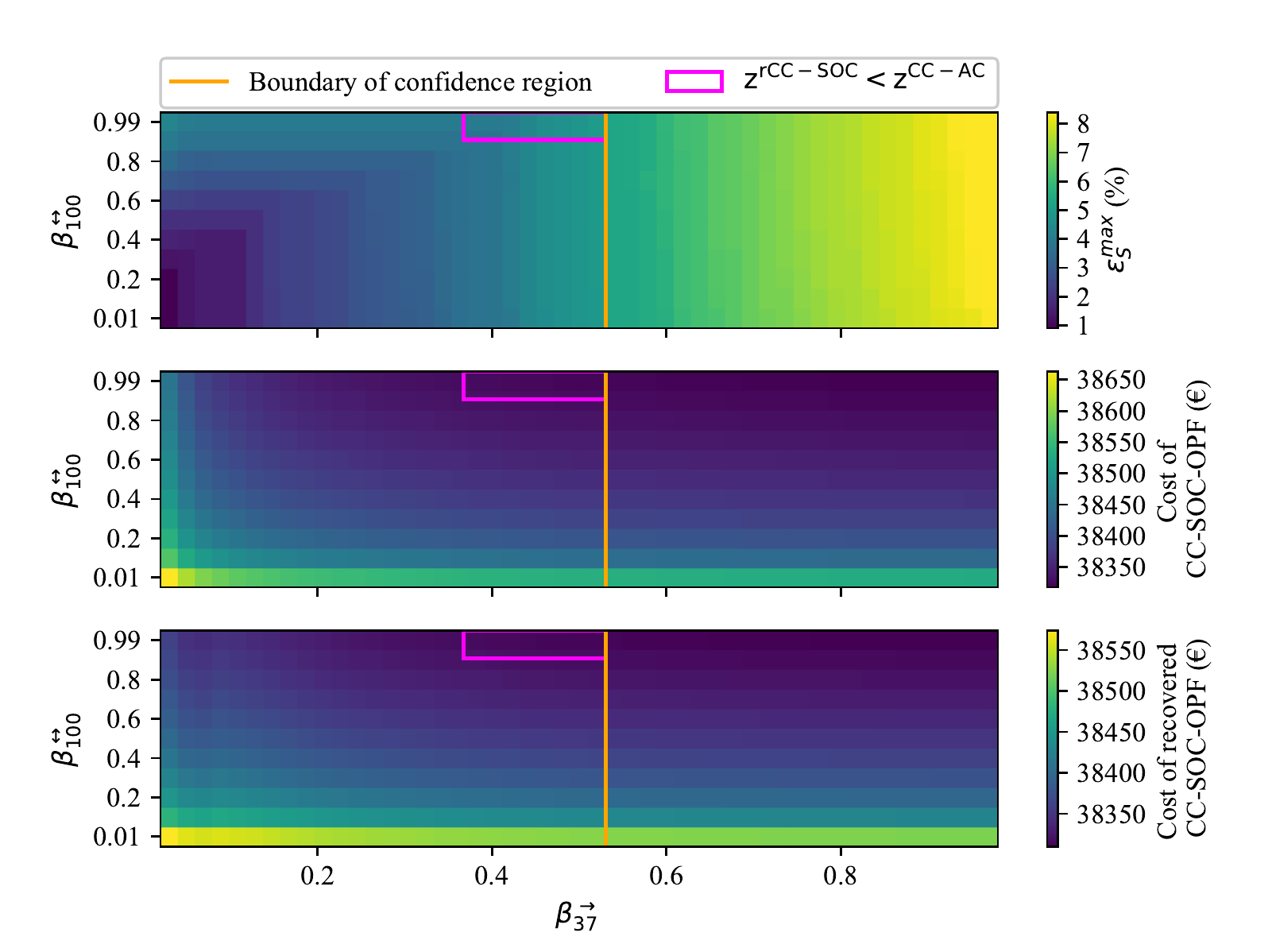}
\centering \caption{Maximum violation probability of apparent branch flows $\epsilon_{S}^{max}$, generation cost of the CC-SOC-OPF ($\textnormal{z}^\textnormal{CC-SOC}$) and the recovered CC-SOC-OPF solution ($\textnormal{z}^\textnormal{rCC-SOC}$) as functions of $\beta^{\rightarrow}_{37}$ and $\beta^{\leftrightarrow}_{100}$. The pink box indicates the region of operating points, which are located inside the CR and are cheaper than the benchmark CC-AC-OPF solution. All operating points left of the boundary are located within the CR.}
\label{fig:allPlot}
\end{figure}

The performance of the resulting 539 operating points when subjected to wind infeed variations is evaluated through Monte Carlo simulations based on 2'000 samples drawn from a Gaussian distribution. The results are visualized in Fig. \ref{fig:allPlot}. The top plot shows the maximum violation probability of the apparent branch flow constraints, which in all 539 cases is due to the flow on line 37. It can be seen that lower values of $\beta^{\rightarrow}_{37}$ and $\beta^{\leftrightarrow}_{100}$ increase the level of conservatism and reduce the violation probability. $\beta^{\rightarrow}_{37}$ needs to be lower than approximately 0.6 to keep the violation probability within acceptable levels (i.e. $<5\%$). We can also observe that for $\beta^{\rightarrow}_{37}>0.4$, variations in $\beta^{\leftrightarrow}_{100}$ do not significantly influence the maximum violation probability. The middle and bottom plots depict the changes in generation cost of the CC-SOC-OPF $z^{CC-SOC}$ and the recovered solution $z^{rCC-SOC}$, respectively. The behavior of the cost development in both cases is similar and leads to an increase in cost with lower weights. Somewhat counter-intuitive though, the cost of the recovered AC feasible solution $z^{rCC-SOC}$ is lower than the cost of the CC-SOC-OPF $z^{CC-SOC}$. This is a consequence of the approximation of the chance constraints in order to make them tractable. As shown in the illustration of ``cut 2'' in Fig.~\ref{fig:FeasibleSpaceCCSOC}, the linear cuts lead to some parts of the AC feasible space being cut off and not represented in the CC-SOC-OPF. Our feasibility recovery procedure however, not constrained by those linear cuts, is able to determine solutions inside the confidence region, which can have lower costs.

\begin{figure}
\centering
\includegraphics[width=3.5in]{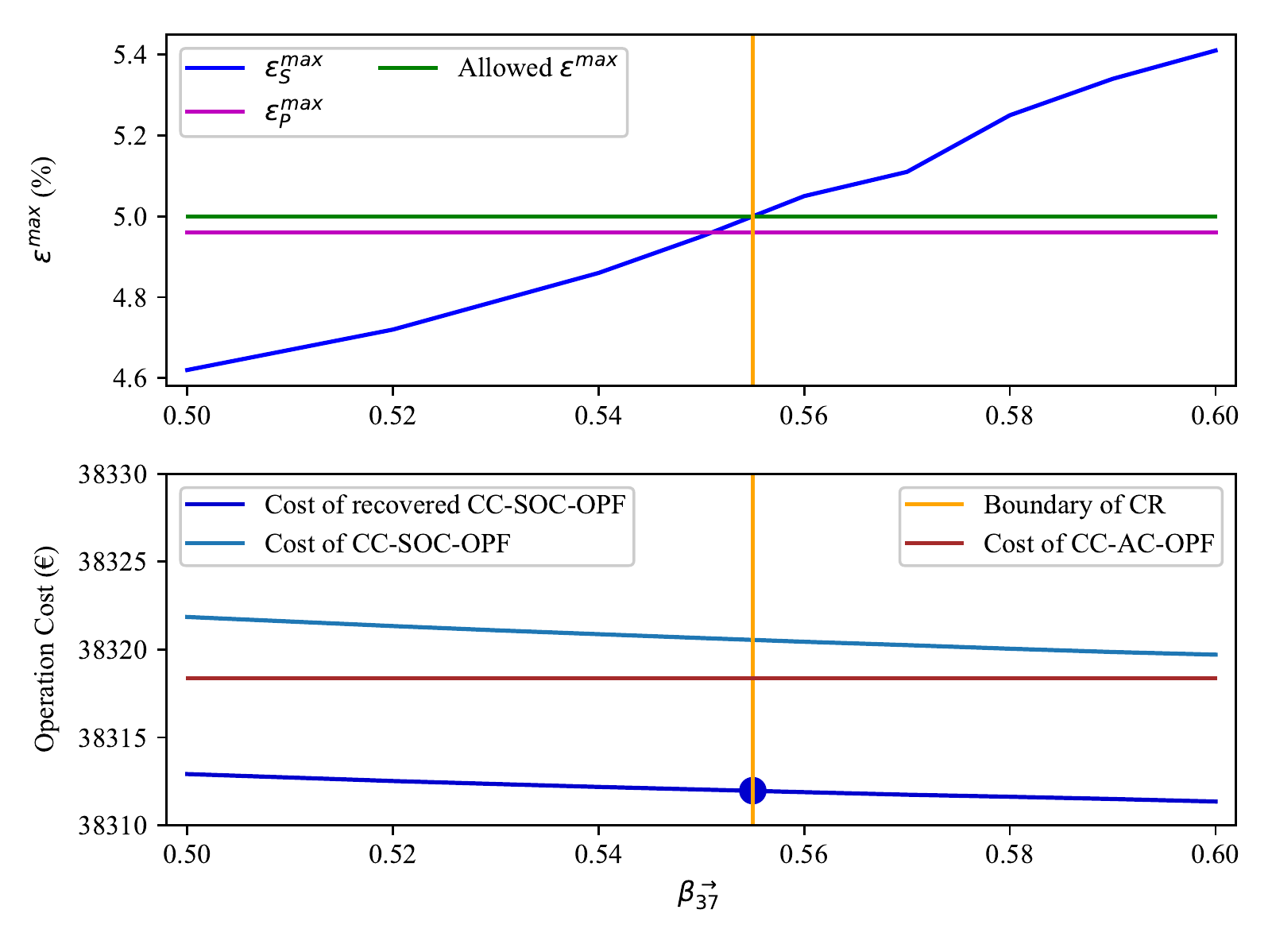}
\centering \caption{Operation cost and maximum violation probability of apparent branch flow $\epsilon_{S}^{max}$ and active power generation $\epsilon_{P}^{max}$ limits for different values of $\beta_{37}^{\rightarrow}$ along line 37 and a constant confidence level $(1-\epsilon)$ for the quadratic chance constraints associated with line 100.}
\label{fig:FineSampling}
\end{figure}

We use a finer sampling of $\beta^{\rightarrow}_{37}$ between 0.5 and 0.6 and evaluate the resulting operating points with 10'000 Monte Carlo simulations to determine its optimal value, which complies with the maximum violation probability but does not lead to unnecessarily high levels of conservatism and cost. We do not assume any weights for the chance constraints associated with line 100, as they are already met when enforced with the usual confidence level. Fig. \ref{fig:FineSampling} depicts the change in violation probabilities and cost for the finer sampling. A value of 0.555 for $\beta^{\rightarrow}_{37}$ has proven to just meet the maximum allowable 5\% violation probability while still leading to lower operation cost at its recovered solution than the CC-AC-OPF. {\color{rev1}Note that the sensitivity analysis has only been performed for the CC-SOC-OPF, as the quadratic chance constraints can only be applied to convex quadratic constraints and are not used within the CC-AC-OPF, where the apparent power flow limits are nonconvex. In the CC-AC-OPF linear sensitivities of the apparent branch flows are used to compute uncertainty margins for the apparent branch flow limits. Their derivation can be found in \cite{Jeremias_SemesterThesis}.}
\subsubsection{Comparison with CC-AC-OPF}
\begin{figure}
\centering
\includegraphics[scale=0.52]{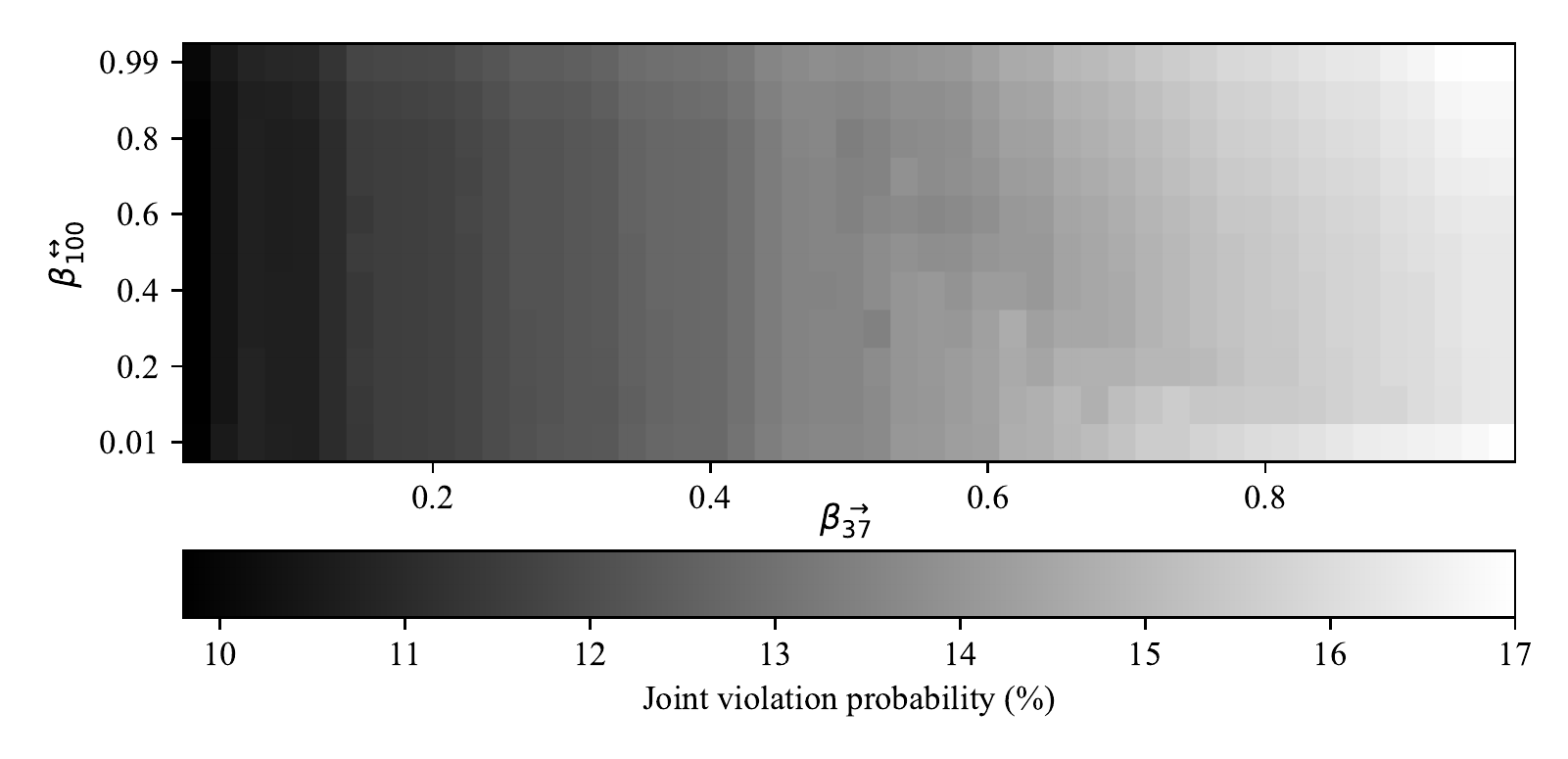}
\centering \caption{Joint violation probability for different values of $\beta_{37}^{\rightarrow}$ and $\beta_{100}^{\leftrightarrow}$ .}
\label{fig:JointViolation}
\end{figure}

The CC-SOC-OPF coupled with the AC feasibility recovery results in an operating point with lower cost as shown in Table \ref{tab:2} and is thus, less conservative. This is also reflected in less conservative violation probabilities shown in Table \ref{tab:1}.
The weights $\beta$ on the quadratic chance constraints provide a significant additional degree of freedom in the CC-SOC-OPF, which can be used to, e.g., reduce the joint violation probability of the original OPF and increase its robustness without the need to explicitly account for joint chance constraints and use computationally demanding sample-based scenario approaches to reformulate them. In this case study, joint violation probabilities of less than 10\% can be achieved as depicted in Fig. \ref{fig:JointViolation}.
\begin{table}
  \caption{Comparison of CC-AC-OPF and CC-SOC-OPF*.}
  \label{tab:2}
  \centering
  \begin{tabular}{lcc}
  \toprule
  & \textbf{CC-AC-OPF} & \textbf{CC-SOC-OPF*} \\ \midrule
  \textbf{Cost} & 38'318.35 \euro & 38'311.12 \euro \\
  \textbf{Iterations} & 4 & 4 \\
  \textbf{Time} & 4.42 s & 10.60 s \\
  \bottomrule
  \end{tabular} \\
  \vspace{0.5ex}
  \centering \scriptsize * Refers to the recovered SOC solution.
\end{table}

The pink box in Fig. \ref{fig:allPlot} highlights the recovered solutions of the CC-SOC-OPF, which are located within the CR and have lower generation cost $z^{rCC-SOC}$ than the CC-AC-OPF $z^{CC-AC}$.
Apart from the operating points depicted in the pink box, 10 other points identified during the finer sampling of $\beta_{37}^{\rightarrow}$ (i.e., $0.5 \leq \beta_{37}^{\rightarrow} \leq 0.6$) also fulfill the original chance constraints and outperform the CC-AC-OPF in terms of operation cost. Thus, apart from the least-cost solution listed in Table \ref{tab:1} and \ref{tab:2}, we find 18 other operating points, which are AC feasible, fulfill the original chance constraints and are still cheaper than the CC-AC-OPF solution of the nonlinear solver. This highlights (i) the potential of convex relaxations to determine the boundaries of the CR and the true optimal of nonconvex problems, and (ii) the importance of appropriate back-mapping procedures to translate the solution of the convex approximation back to the original domain. Despite the required tuning, $\beta$ provides the flexibility to vary the shape of the convex approximation and direct the solution from a true lower bound back into the original feasible space and the CR.

The need for computationally more efficient convex relaxations of (chance-constrained) AC-OPFs was identified in \cite{Andreas_CCACOPF}, where we developed a SDP relaxation of the chance-constrained AC-OPF based on rectangular and Gaussian uncertainty sets. Comparable instances to our case study may take up to 10 minutes to solve with the SDP relaxation (although the computational improvements proposed in \cite{Venzke_SDP_SCOPF_PSCC2018} can reduce this time) whereas our proposed algorithm converges within 10.60 s. The solution time of the CC-SOC-OPF is mainly determined by the inner iteration loop for approximating the angle constraint \eqref{eq:ATAN2}, which accounts for 84\% of the total solution time. More efficient approximations of the angle constraint, which can be implemented in an one-shot optimization, could significantly improve the performance of the proposed method. The CC-AC-OPF converges even faster after only 4.42 s{\color{rev1}, which again demonstrates its efficiency, when it converges. However, the discussion in Section \ref{sec:robustness_extensions} highlighted the need for more robust solution techniques not only for the CC-AC-OPF but also for the standard AC-OPF \cite{Anders_RobustnessAndScalabilityOfSDP}, which needs to be considered when comparing the two solution approaches. Another example is the work in \cite{ferc_SolutionTechniquesACOPF}, where the authors compared the performance of different solution techniques for the nonlinear AC-OPF. They demonstrated how the convergence behavior and solution quality in terms of costs for both small and large networks was highly sensitive to (i) the initialization of the various tested solvers (i.e., warm start) and (ii) the OPF problem formulation (i.e., rectangular, polar etc.). As a consequence, the authors strongly recommended not to rely on one solution technique only, but to employ a \textit{multistart strategy} in real-life networks, where several solution techniques with different solver intializations are run in parallel to increase the robustness of the AC-OPF solution.
In view of this, a \textit{multistart strategy} could also be employed in case of the chance-constrained AC-OPF, where various instances of both the CC-SOC-OPF and the CC-AC-OPF are run in parallel to ensure robustness for the convergence and the lowest system cost. 
}

Still, the combination of the iterative algorithm and convex programming makes the method computationally very efficient, robust, and suitable for large-scale systems as demonstrated in our case study. Most current industrial tools integrate OPF calculations in iterative frameworks along with other functionalities (e.g., security assessments) \cite{opf_basic_requirements}. Consequently, the iterative solution algorithm with the decoupled uncertainty assessment is well aligned with this framework and has significant potential for application in already existing calculation procedures \cite{Line_CCACOPF}. {\color{rev2} However, it must be noted that in case of infeasibility, our approach relies on the availability of a robust AC power flow tool to ensure a reliable AC feasibility recovery. AC power flow algorithms are usually based on an iterative numerical technique for solving a set of nonlinear equations and their convergence depends on an appropriate initialization. Nevertheless, we expect the solution of the CC-SOC-OPF to be a good initial guess, while the AC power flow algorithms are at a mature development stage. As a result, convergence issues, if any, are expected to be rare.  
}

\section{Conclusion} \label{sec:Conclusion}
This paper deals with the impact of linear approximations for the unknown nonconvex confidence region of chance-constrained AC-OPF problems. 

In that, we introduce the first formulation of a chance-constrained second-order cone OPF. Our approach is superior to existing approaches, as it defines a convex problem and thus, provides convergence guarantees, while it is computationally more efficient than other convex relaxation approaches. Coupled with an AC feasibility recovery, we show that it can determine better solutions than chance-constrained nonconvex AC-OPF formulations {\color{rev1}and is guaranteed to provide a solution even in cases, where nonlinear solvers already fail for the standard deterministic AC-OPF \cite{Anders_RobustnessAndScalabilityOfSDP}}.

To the best of our knowledge, this is the first paper that performs a rigorous analysis of the AC feasibility recovery for robust SOC-OPF formulations. Due to the SOC relaxation, a CC-SOC-OPF might determine AC infeasible operating points, while the linear reformulations of the chance constraints result in solutions, which either lie outside the confidence region or are too conservative. Inaccurate approximations of the confidence region is an issue for all chance-constrained AC-OPF formulations. In this paper, we introduce an approximation of the quadratic apparent power flow chance constraints with linear chance constraints using results proposed in \cite{TwoSided_LinCC}. Through that, we introduce new parameters able to reshape the approximation of the confidence region, offering a high degree of flexibility.
{\color{rev1}
\section{Future Work} \label{sec:FutureWork}}
Our paper shows that further work on better and computationally more efficient approximations for the chance-constrained AC-OPF problem is necessary.

{\color{rev1}The major challenge of the CC-SOC-OPF lies in the optimal selection of $\beta$ for approximating the quadratic chance constraints. While we have proposed an ex ante screening to reduce the effort associated with the parameter tuning, more systematic procedures are necessary to ensure an optimal setting for realistic systems. One potential solution approach could be derived from the recent work in \cite{Xie_OptimizedBonferroni}, which defines a framework for optimizing the Bonferroni approximation, where the violation probabilities, which are aligned with $\beta$ in our setting, are optimization variables and not known a priori.

Other possible directions for extensions include distributionally robust formulations for the CC-SOC-OPF and the consideration of integer variables and large uncertainty ranges in the proposed framework.

Furthermore, we are planning to use a combination of data-driven methods from our work in \cite{DataDrivenSCOPF,DataDriven_SOCOPF_PSCC2018}, convex relaxations, and the iterative solution framework to develop a scalable approach to {an \color{rev1}integrated security-} and chance-constrained OPF. In \cite{DataDrivenSCOPF,DataDriven_SOCOPF_PSCC2018} we propose a novel approach which efficiently incorporates N-1 and stability considerations in an optimization framework and is suitable for integration in the proposed CC-SOC-OPF framework.

}

{\color{rev1}
\appendix
\subsection{Derivation of the Linear Sensitivities}
The linear sensitivity factors are calculated at each iteration of the CC-SOC-OPF based on a linearization around the iteration's current optimal operating point $\mathbf{y^*}$. The changes in nodal active and reactive power injections can be expressed in terms of the wind deviation $\mathbf{\xi}$, the generator participation factors $\mathbf{\gamma}$, the unknown nonlinear changes in active and reactive power (i.e., $\mathbf{\Delta P^{U}}$ and $\mathbf{\Delta Q}$) and the ratio $\mathbf{\lambda}$ between the reactive and active power injection of wind farms. 
Thus, the left-hand side of \eqref{eq:JACSOC} can also be expressed as follows:
\begin{gather}
\begin{bmatrix}
\mathbf{-\gamma 1} \\
\mathbf{Z} \\
\mathbf{Z} \\
\mathbf{Z}
\end{bmatrix} \mathbf{\xi} +
\begin{bmatrix}
\mathbf{\Delta P^{U}} \\
\mathbf{\Delta Q} \\
\mathbf{0} \\
\mathbf{0}
\end{bmatrix} +
\begin{bmatrix}
\mathbf{I} \\
\mathbf{diag(\lambda)} \\
\mathbf{Z} \\
\mathbf{Z}
\end{bmatrix} \mathbf{\xi} =
\begin{bmatrix}
\mathbf{\Delta P^U} \\
\mathbf{\Delta Q} \\
\mathbf{0} \\
\mathbf{0}
\end{bmatrix} +
\begin{bmatrix}
\mathbf{\Psi}
\end{bmatrix} \mathbf{\xi}, \label{eq:LHS_JACSOC}
\end{gather}
where $\mathbf{Z}$, $\mathbf{1}$ and $\mathbf{I}$ denote $(\lvert\mathcal{N}\rvert \times \lvert\mathcal{W}\rvert)$ or $(\lvert\mathcal{L}\rvert \times \lvert\mathcal{W}\rvert)$ zero, all-ones and identity matrices, respectively. $\mathbf{0}$ is a vector of zeros. The system of equations \eqref{eq:JACSOC} can finally be reformulated to:
\begin{gather}
\begin{bmatrix}
\mathbf{\Delta P^U} \\
\mathbf{\Delta Q} \\
\mathbf{0} \\
\mathbf{0}
\end{bmatrix} +
\begin{bmatrix}
\mathbf{\Psi}
\end{bmatrix} \mathbf{\xi} =
\begin{bmatrix}
\mathbf{J^{SOC}}
\end{bmatrix} \Bigg\rvert_{\mathbf{y^{*}}}
\begin{bmatrix}
\mathbf{\Delta u} \\
\mathbf{\Delta c} \\
\mathbf{\Delta s} \\
\mathbf{\Delta \theta}
\end{bmatrix}. \label{eq:DerivationSensitivity1}
\end{gather}
$\mathbf{\Delta P^U}$ refers to the unknown changes in nonlinear active power losses, which are not accounted for by the generator participation factors.
Following the assumptions outlined in Section~\ref{sec:LDR}, the nonzero elements of $\mathbf{\Delta P^{U}}$ and $\mathbf{\Delta Q}$ are summarized in $\mathbf{\Delta g} := [\Delta P^{U}_{ref} \; \Delta Q_{ref} \; (\mathbf{\Delta Q_{PV}})^{^\mathbf{T}}]^{\mathbf{T}}$. Similarly, $\mathbf{\Delta \hat{y}}$ denotes the nonzero changes in the right-hand side of \eqref{eq:DerivationSensitivity1} (i.e., $\mathbf{\Delta \hat{y}} := [\mathbf{\Delta u_{PQ}^T} \; \mathbf{\Delta c^T} \; \mathbf{\Delta s^T} \; \mathbf{\Delta \theta_{PV}^{T}} \; \mathbf{\Delta \theta_{PQ}^{T}}]^{\mathbf{T}}$). Rearranging \eqref{eq:DerivationSensitivity1} by grouping the nonzero and zero elements separately, i.e.,
\begin{gather}
\begin{bmatrix}
\mathbf{\Delta g} \\
\mathbf{0}
\end{bmatrix} =
\begin{bmatrix}
\mathbf{J_{x}^{SOC,\Romannum{1}}} & \mathbf{J_{x}^{SOC,\Romannum{2}}} \\
\mathbf{J_{x}^{SOC,\Romannum{3}}} & \mathbf{J_{x}^{SOC,\Romannum{4}}}
\end{bmatrix}
\begin{bmatrix}
\mathbf{0} \\
\mathbf{\Delta \hat{y}}
\end{bmatrix} -
\begin{bmatrix}
\mathbf{\Psi_x^{\Romannum{1}}} \\
\mathbf{\Psi_x^{\Romannum{2}}}
\end{bmatrix} \mathbf{\xi}, \label{eq:DerivationSensitivity2}
\end{gather}
allows us to derive linear relationships between the changes in the variables of interest and the wind deviation $\mathbf{\xi}$,
\begin{gather}
\mathbf{\Delta \hat{y}} = \Big( \mathbf{J_{x}^{SOC,\Romannum{4}}}\Big)^{-1} \mathbf{\Psi_{x}^{\Romannum{2}}} \mathbf{\xi} = \mathbf{\Upsilon_{\hat{y}}} \mathbf{\xi}, \label{eq:DerivationSensitivity3} \\
\mathbf{\Delta g} =  \Big( \mathbf{J_{x}^{SOC,\Romannum{2}}} (\mathbf{J_{x}^{SOC,\Romannum{4}}})^{-1}
\mathbf{\Psi_x^{\Romannum{2}}} - \mathbf{\Psi_x^{\Romannum{1}}} \Big) \mathbf{\xi} = \mathbf{\Upsilon_g} \mathbf{\xi}. \label{eq:DerivationSensitivity4}
\end{gather}
Subscript $\mathbf{x}$ in $\mathbf{J_{x}^{SOC}}$ and $\mathbf{\Psi_{x}}$ denotes that the columns and/or rows of the original matrices have been rearranged according to the grouping of zero and nonzero elements. The linear sensitivity factors $\mathbf{\Upsilon}$ are then used to calculate the uncertainty margins $\mathbf{\Omega}$.
}

\ifCLASSOPTIONcaptionsoff
  \newpage
\fi
%
\bibliographystyle{IEEEtran}
\bibliography{library}

\newpage

\begin{IEEEbiography}[{\includegraphics[width=1in,height=1.25in,clip,keepaspectratio]{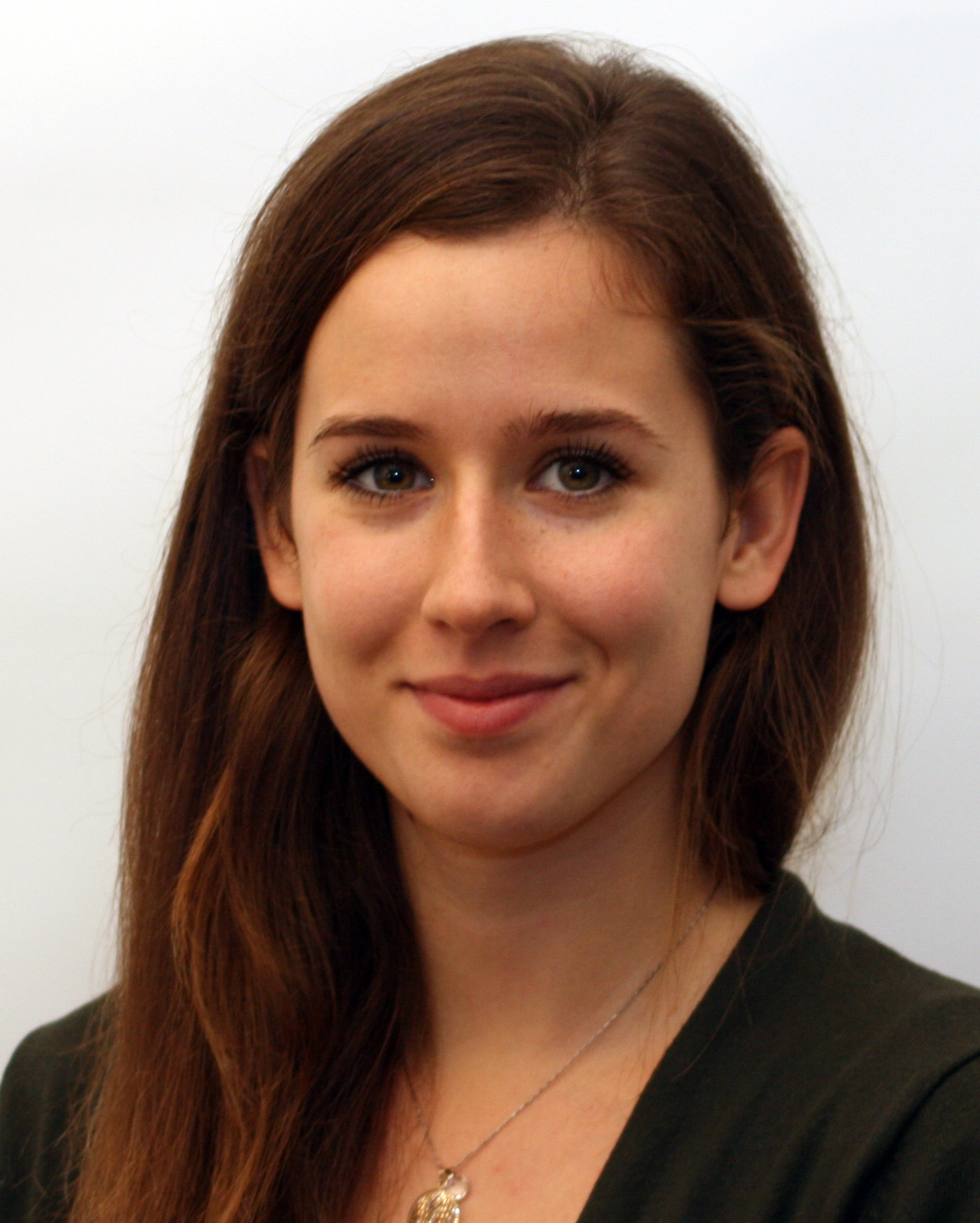}}]{Lejla Halilba\v{s}i\'{c}} (S'15) received the M.Sc. degree in Electrical Engineering from the Technical University of Graz, Austria in 2015. She is currently working towards the PhD degree at the Department of Electrical Engineering, Technical University of Denmark (DTU). Her research focuses on optimization for power system operations including convex relaxations, security-constrained optimal power flow and optimization under uncertainty. 
\end{IEEEbiography}
\begin{IEEEbiography}[{\includegraphics[width=1in,height=1.25in,clip,keepaspectratio]{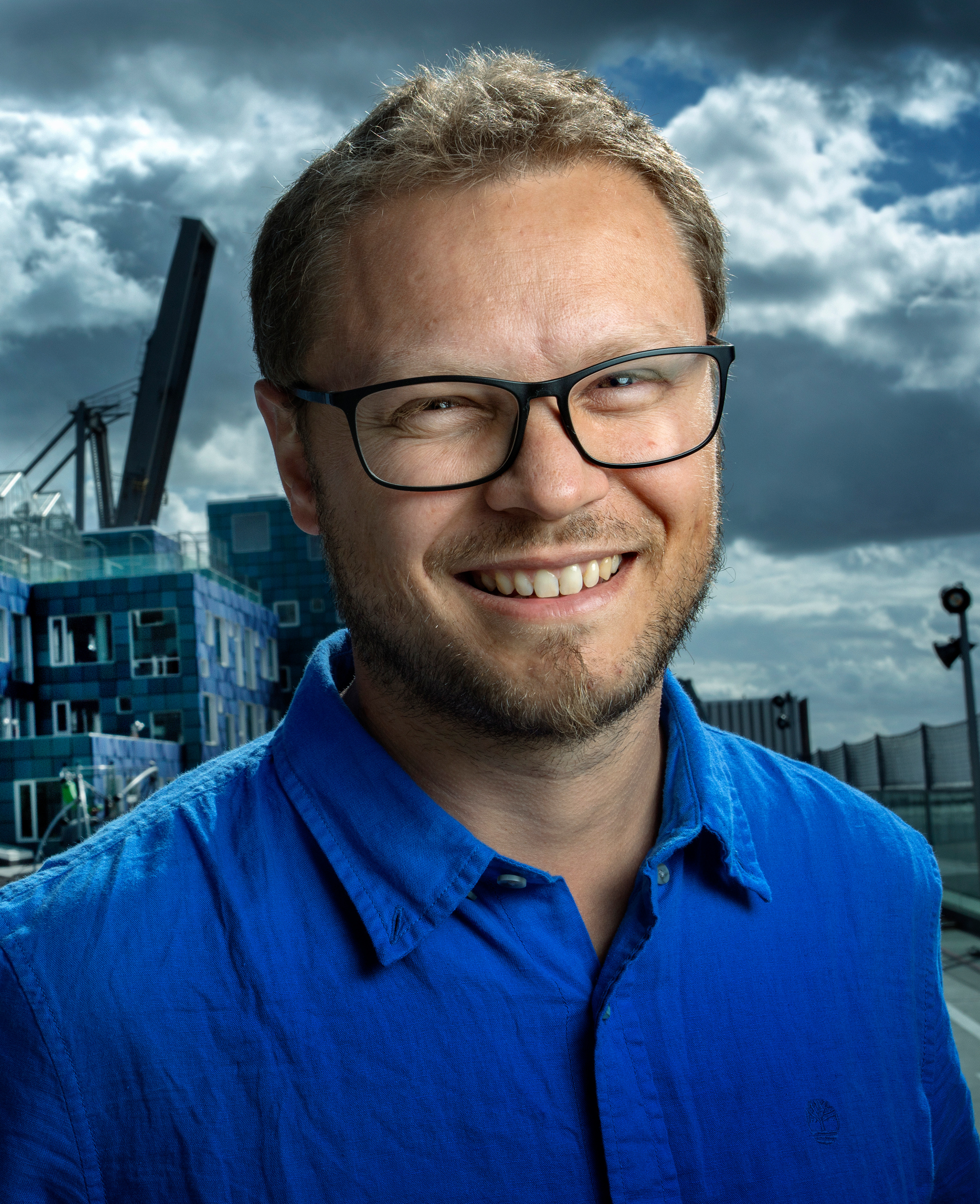}}]{Pierre Pinson} (M'11, SM'13) received the M.Sc. degree in applied mathematics from the National Institute for Applied Sciences (INSA Toulouse, France) and the Ph.D. degree in energetics from Ecole des Mines de Paris (France). He is a Professor at the Technical University of Denmark (DTU), Centre for Electric Power and Energy, Department of Electrical Engineering, also heading a group focusing on Energy Analytics \& Markets. His research interests include among others forecasting, uncertainty estimation, optimization under uncertainty, decision sciences, and renewable energies. 
Prof. Pinson acts as an Editor for the International Journal of Forecasting, and for Wind Energy.
\end{IEEEbiography}
\begin{IEEEbiography}[{\includegraphics[width=1in,height=1.25in,clip,keepaspectratio]{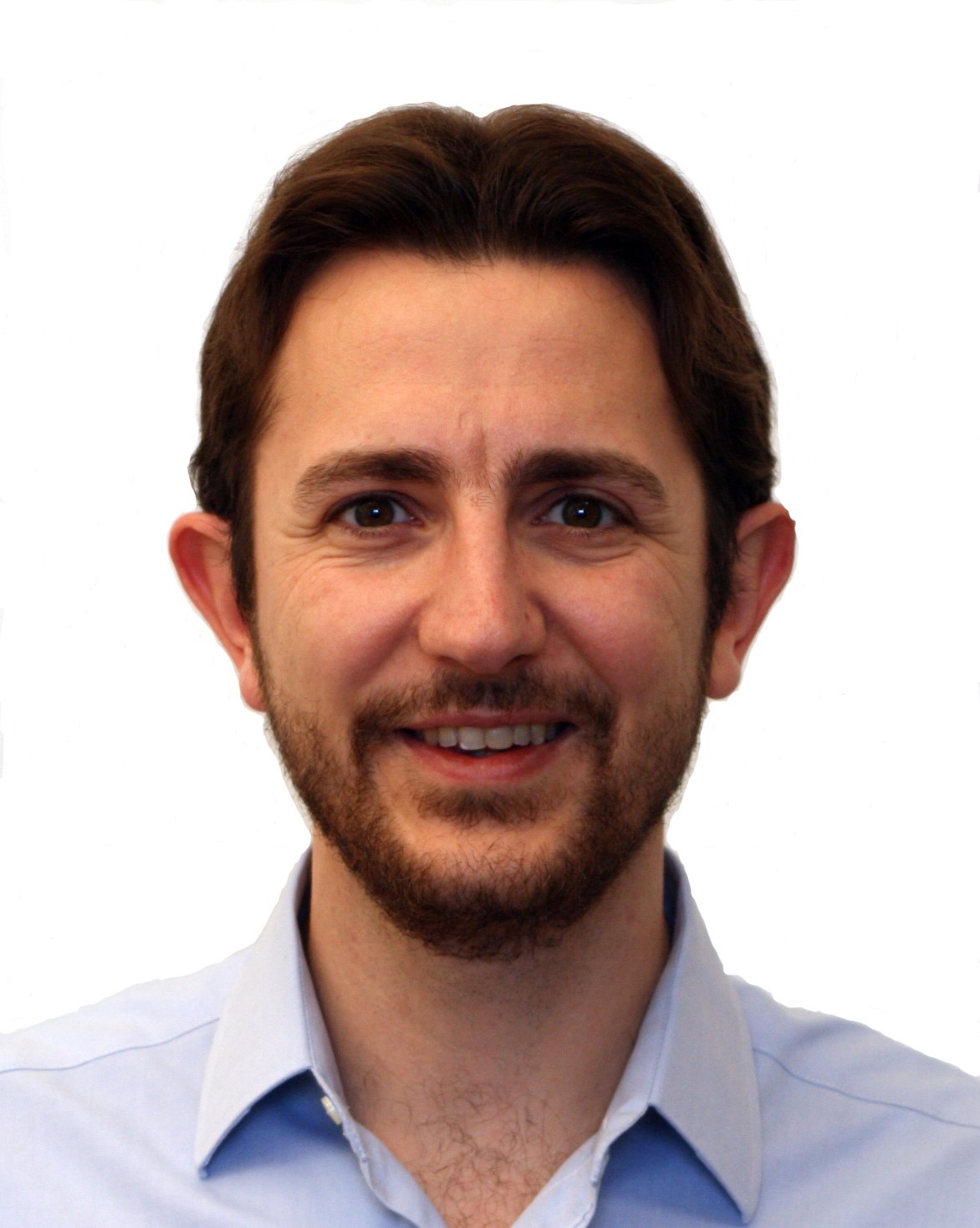}}]{Spyros Chatzivasileiadis} (S'04, M'14, SM'18) is an Associate Professor at the Technical University of Denmark (DTU). Before that he was a postdoctoral researcher at the Massachusetts Institute of Technology (MIT), USA and at Lawrence Berkeley National Laboratory, USA. Spyros holds a PhD from ETH Zurich, Switzerland (2013) and a Diploma in Electrical and Computer Engineering from the National Technical University of Athens (NTUA), Greece (2007). In March 2016 he joined the Center of Electric Power and Energy at DTU. He is currently working on power system optimization and control of AC and HVDC grids, including semidefinite relaxations, distributed optimization, and data-driven stability assessment.    
\end{IEEEbiography}

\vfill
\end{document}